\documentclass[%
 reprint,
 amsmath,amssymb,
 aps,
]{revtex4-2}

\usepackage{graphicx}% Include figure files
\usepackage{dcolumn}% Align table columns on decimal point
\usepackage{bm}% bold math
\usepackage{xcolor}%color blocks of text
\usepackage{float}
\usepackage{placeins}
\usepackage[normalem]{ulem}
\usepackage{soul}
\newcommand{\vR}{\mathcal{R}}
\newcommand{\vL}{\mathcal{L}}

\begin{document}

\preprint{APS/123-QED}

%\title{Condensin forms catch bonds}% Force line breaks with \\

%\title{Loop Extrusion Reversal by Condensin involves Catch Bonds}
\title{Loop Extrusion Reversal by Condensin Motor is Mediated by Catch Bonds}
\author{Atreya Dey}
\affiliation{Department of Chemistry, The University of Texas at Austin, Austin, Texas 78712, USA}

\author{Guang Shi}
\affiliation{Department of Chemistry, The University of Texas at Austin, Austin, Texas 78712, USA}

\author{Ryota Takaki}
\affiliation{Department of Chemistry, The University of Texas at Austin, Austin, Texas 78712, USA}

\author{D. Thirumalai}
\email{Dave.Thirumalai@gmail.com}
\affiliation{Department of Chemistry, The University of Texas at Austin, Austin, Texas 78712, USA}
\affiliation{Department of Physics, The University of Texas at Austin, Austin, Texas 78712, USA}
% \affiliation{%
%  Authors' institution and/or address\\
%  This line break forced with \textbackslash\textbackslash
% }%

%\date{\today}

\begin{abstract}
Structural Maintenance Complexes (SMC) are energy consuming motors that are important in folding the genome  by loop extrusion (LE) in all stages of the cell cycle. Single molecule magnetic tweezer pulling experiments have revealed that condensin, a member of the SMC family involved in mitosis, takes occasional backward steps, thus coughing up the gains in the length of the extruded loop. To reveal the mechanism of the forward and backward steps simultaneously, we developed a theory using the stochastic kinetic model and the scrunching mechanism for LE. The calculations quantitatively account for the measured force-dependent step size and dwell time distributions in both the directions. By postulating the existence of an intermediate state in the ATP-driven cycle that is poised to take a forward or a backward step, we predict that its lifetime increases as the external mechanical force increases till a critical value and subsequently decreases at higher forces.   The surprising finding of lifetime increase in an active motor, at sub-piconewton forces, is the characteristic of catch bonds, known in force-induced rupture of several passive protein complexes. The identification of catch bond-like states in condensin not only expands our understanding of LE but also highlights the significance of mechanical forces in regulating genome organization.
%Condensin is a well-known loop extruding molecule that plays a crucial role in chromosome organization and gene regulation. Recent experimental studies have revealed that condensin can take occasional backward steps. In this study, we extend our previous scrunching mechanism of loop-extrusion and propose a two-state model. We demonstrate that the lifetime of one of these states transiently increases in response to external forces with a peak around 0.4 pN, implying catch-bond like behavior. By performing Gillespie simulations with our two-state model, we predict that loop extrusion by condensin could resonate with an external mechanical  force. Finally, we propose experimental strategies to verify our predictions. Our findings shed light on the intricate kinetics of condensin and provide insights into its ability to dynamically regulate chromosome structure. The identification of catch bond-like microstates and their force response not only expands our understanding of loop extrusion processes but also highlights the significance of mechanical forces and noise in biological systems.
% \begin{description}
% Secondary publications and information retrieval purposes.
% You may use the \texttt{description} environment to structure your abstract;
% use the optional argument of the \verb+\item+ command to give the category of each item. 
% \end{description}
\end{abstract}

\maketitle

\section{Introduction}
The family of molecular motors, referred to as structural maintenance complexes (SMCs) \cite{uhlmann2016smc,nasmyth2005structure}, have evolved to fold chromosomes, apparently without knots~\cite{goundaroulisBiophysicalJournal2020}, so that they can be housed in the small volume of the nucleus. The SMC motors such as condensins I and II~\cite{Hirano1997cell}, cohesin
~\cite{Michaelis1997cell}, and SMC 5/6~\cite{pradhan2023smc5} facilitate chromosome folding by undergoing allosteric transitions between various states during the catalytic cycle of the motor~\cite{Takaki21NatComm,Marko19NAR}. The dynamical allosteric transitions are driven by ATP binding, resulting in the extrusion of chromatin loops \cite{hassler2018towards,Takaki21NatComm,dekker2023molecular}. The condensation of a large array of overlapping loops of varying sizes by ATP-dependent SMC motors eventually results in the folding of the genome~\cite{Dekker23Science}. Single molecule \textit{in vitro} experiments have been used to directly visualize the process of loop extrusion (LE) \cite{ganji2018real,pradhan2023smc5,davidson2019dna,kim2019human,pobegalov2023single}. A few mechanisms have been proposed to describe loop extrusion, sometimes without treating SMCs as energy consuming motors~\cite{Aiipour12NAR,Fudenberg16CellRep}.
We prefer  the ``scrunching" mechanism~\cite{ganji2018real,Takaki21NatComm}, initially proposed in the context of bubble formation in DNA during bacterial transcription~\cite{Revyakin06Science,Kapanidis06Science}. The theory  proposed using the scrunching mechanism quantitatively  explained the measured force-velocity relation~\cite{ganji2018real} and the step size distribution as a function of the external force. 

The major results of the \textit{in vitro} experiments are: (1) By consuming no more than two ATP molecules during a single catalytic cycle, loops are extruded  at a velocity ($\Omega$) that is on the order of 1,000 - 1,500 $\text{bp/s} \approx 510\ \text{nm/s}$ in the absence of any external load~\cite{ganji2018real}.  The velocity decreases continuously with load and approaches zero at the stall force $\approx 0.8$ pN~\cite{ganji2018real}. (2) The mean loop length extruded per step is on the order of the size of the motor ($\sim 50$ nm). Using $\Omega$=1,500 $\text{bp/s}$ and the lifetime of condensin bound to DNA to be 120$s$~\cite{Dey23CellReports} and the mean 
extrusion step size to be $\sim 50$ nm, can estimate that the motor takes $\approx$ 140 steps before disengaging from DNA.   (3) Strikingly, the distribution of loop lengths extruded in a single step is broad~\cite{ganji2018real,pradhan2023smc5,davidson2019dna,kim2019human,golfier2020cohesin}. Although the mean step size is $\sim 50$ nm, condensin takes unusually large steps ($\sim (100$ - $150$) nm or $\sim (300 - 450)$ bp) with non-negligible probability, far exceeding the size of the motor. 
(4) Finally,  as required by the principle of microscopic reversibility~\cite{tolman1925principle} condensins (and presumably cohesin) motors could take ``backward" steps just like conventional motors (kinesin, myosin, and dynein)~\cite{Mugnai20RMP,kolomeisky2007molecular}. Indeed, recently magnetic tweezer experiments were used to measure~\cite{ryu2022condensin} the number of forward and backward (reverse) steps taken by condensin along with the corresponding dwell time distributions. It was found that when condensin takes a backward step, a portion of the extruded loop is lost, which means that the length of the extruded loop decreases from $L$ (before the backward step) to $L - \Delta L$ ($\Delta L > 0$). In the limited number of measured trajectories~\cite{ryu2022condensin}, loop reversal was always preceded by forward steps. This is not  the case in the stepping of conventional motors~\cite{Mugnai20RMP} in which there is no correlation between the probability of backward step (the analogue of loop reversal in condensin) and the nature of the preceding step.    

Although both theory~\cite{Takaki21NatComm,Yamamoto2017pre,Brackley2017prl,Marko19NAR} 
and simulations \cite{pobegalov2023single, Brackley2018nucleus, Bonato2021biophyjournal, Higashi2021elife}  have been proposed to explain the forward steps, with the exception of our previous study, none of them have reproduced the \textit{in vitro} measurements quantitatively. Recently, we developed a theory \cite{Takaki21NatComm}, using polymer physics and fluctuation theorem to quantitatively explain the first two results summarized above. Our theory, which used the scrunching mechanism as the basis for loop extrusion, quantitatively reproduced the experimental results for extrusion velocity ($\Omega$) as a function of an external mechanical force, $F$, that is applied to DNA. It also predicted $\Omega$ as a function of ATP concentration and $F$. Validation of this prediction as well as the load-dependent step-size distribution await future experiments. Importantly, to our knowledge, it is the only theory~\cite{Takaki21NatComm} that explains the experimental findings for the distribution of the length of extruded loops.

Although the forward stepping is understood theoretically~\cite{Takaki21NatComm} and through simulations, there is no explanation for the observed~\cite{ryu2022condensin} mechanism of occasional backward steps that condensin takes during the loop extrusion process.   Understanding the backward stepping is important because  it will 
further elucidate  the mechanochemical details of the loop extrusion cycle. 
Here, we  develop a stochastic kinetic model to account for  both forward and backward loop steps.   The theory shows that the experimentally observed~\cite{ryu2022condensin}   backward steps may be accounted for by postulating the presence of an intermediate state in the extrusion cycle. The calculated dwell times and the distributions of step sizes in the forward and backward directions are in good agreement with magnetic tweezer experiments. Surprisingly, we find that the lifetime of the intermediate state, which plays an important role in controlling the mechanism of the forward and backward steps,  increases with external load until a critical value and subsequently decreases. The predicted counterintuitive response to force, manifested in the increase followed by the decrease in the lifetime of the intermediate, is reminiscent of catch bonds known in a variety of biological contexts \cite{dembo1988reaction,Barsegov05PNAS,Barsegov06JPCB,Thomas2008ARB,Marshall03Nature,Huang17Science,Luca17Science,Choi25AnnRevImm}.

\section{Theory}
\begin{figure*}[ht]
    \centering
    \includegraphics[width=\linewidth]{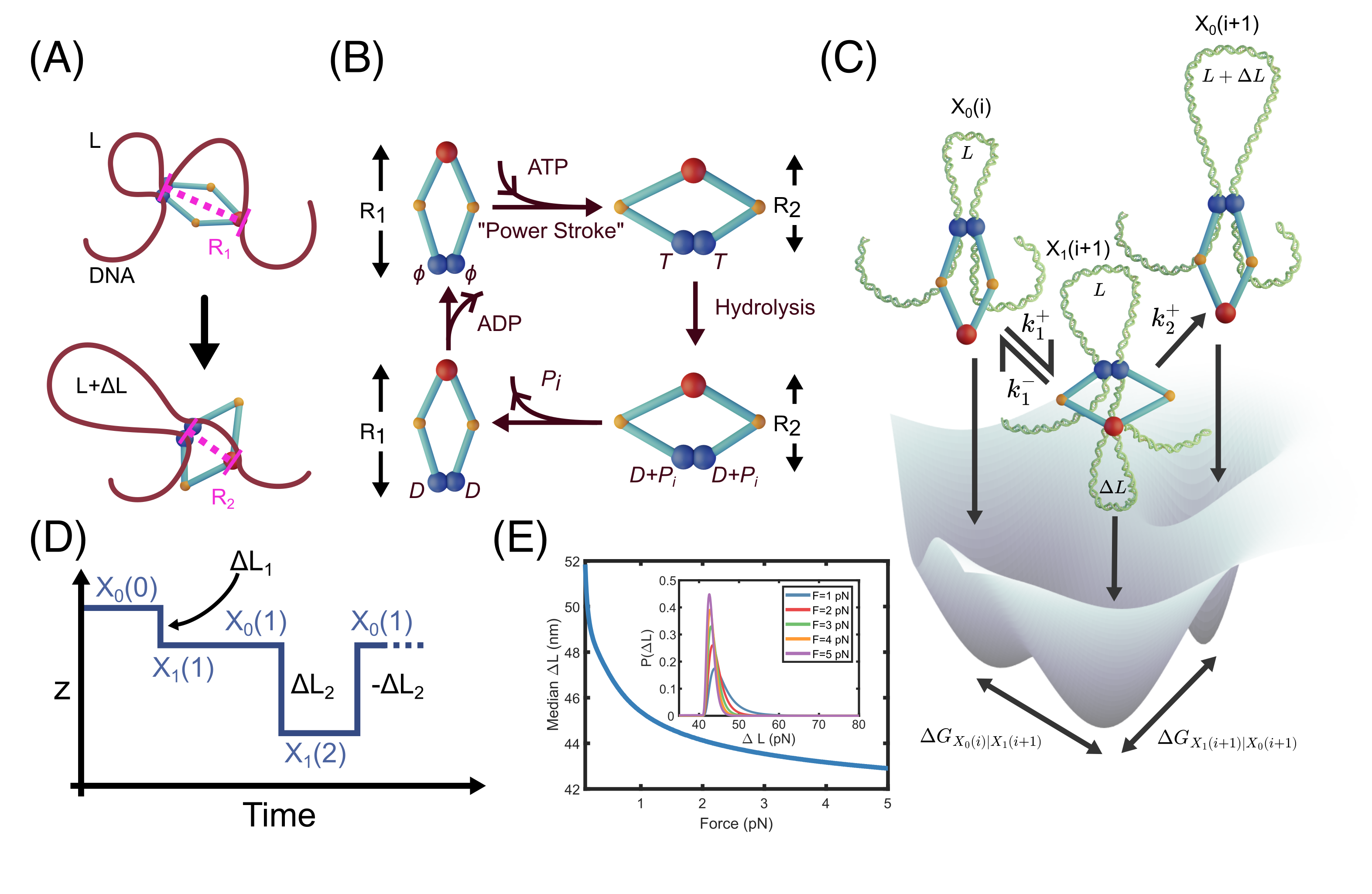}
    \caption{\textbf{Schematic of the catalytic cycle and conceptualization of loop extrusion:}(A) Scrunching mechanism of loop extrusion by SMC in general and condensin in particular in the forward step. Upon ATP binding, the motor 
    head (represented by two blue spheres) undergoes an allosteric transition that brings the hinge (purple sphere) close to the motor heads. In the process, the distance between the head and the hinge decreases from $R_1$ to $R_2$. The loop size increases from $L$ to $L+ \Delta L$. (B) Sketch of the condensin reaction cycle.  ATP binding, that drives the allosteric transition, is the ``power stroke".  (C) Loop extrusion model allowing for both forward and reverse stepping. During loop extrusion, fueled by ATP binding, condensin shuttles between the states $X_0$ and $X_1$. The state $X_0(i+1)$ is equivalent to state $X_0(i)$ for the subsequent step in extrusion. The transition between $X_0(i)$ and $X_1(i+1)$ is modeled as a reversible process, whereas the transition between state $X_1(i+1)$ and $X_0(i+1)$ is assumed to be irreversible. (D) Sketch of trajectory during  of loop extrusion measurable in magnetic tweezer experiments. The x-axis denotes the observation time, while the y-axis denotes the position of the magnetic bead in the z-direction in the experiments \cite{ryu2022condensin}. Negative (positive) $z$ values correspond to forward (backward) steps.  (E) Median $\Delta L$ as a function of external force. The inset shows the distribution,  $P(\Delta L)$,  at various force values.}
    \label{fig:LEmodel}
\end{figure*}

\textbf{Physical picture}: All energy consuming molecular motors~\cite{Block07BJ,Block94Cell,Mugnai20RMP,kolomeisky2007molecular}, including SMCs, undergo a catalytic cycle that can be broken down into the following steps: ATP binding to the \textit{apo} state, ATP hydrolysis, release of the inorganic phosphates and ADP, and finally resetting to the starting state \textit{apo} state. Unlike in the well-studied cargo-carrying linear motors~\cite{Svoboda94ARBB,kolomeisky2007molecular,Holzbaur10COCB,Mugnai20RMP,Isojima16NatChemBiol}, the structural changes in the states (\textit{apo}, ATP-bound, ADP) that are involved in condensin (or cohesin) are not fully characterized. Nevertheless, a  plausible catalytic cycle  \cite{Marko19NAR,Takaki21NatComm} that is useful in creating theories for loop extrusion has been suggested (Fig.~\ref{fig:LEmodel}A). Based on an earlier study~\cite{Takaki21NatComm}, we posit that the power stroke is associated with the binding of ATP to the motor head domain. We proposed that ATP binding triggers an allosteric transition that changes the distance between the hinge 
domain and head domain from $R_1$ to 
$R_2$ ($R_2 < R_1$) by the ``scrunching" mechanism, which was first proposed to explain the creation of the DNA bubble in the early stages of bacterial transcription~\cite{Revyakin06Science,Kapanidis06Science}. In the context of condensin, the same mechanism assumes that the hinge-motor head distance changes during the catalytic cycle while the motor heads, for all practical purposes, are stationary.   After the power stroke,  ATP is hydrolyzed, inorganic $P_i$ and ADP are released resetting in the motor to the \textit{apo} state. Subsequently, ATP binds to the heads, thus restoring the motor to the initial state, poising the motor to begin a new cycle.

The transitions between the allosteric states traversed by the motor enables condensin and cohesin to extrude loops of varying lengths. We envision that the process occurs, as sketched in Fig.~\ref{fig:LEmodel}. The state of condensin is written as $X_0(i)$ or $X_1(i)$ where $i$ is the index for extrusion steps and the subscripts $0$ or $1$ denote whether condensin is the O-state or the B-state~\cite{ryu2020condensin}. At the start of the extrusion process, the motor is in a state labeled $X_0(i)$  containing an extruded  loop of length $L$, which is allowed to be zero.  Transitions $X_0(i)\rightarrow X_1(i+1)$ occurs at a rate $k_1^+$, and the rate for the reverse process is $k_1^-$. If there is a transition to $X_0(i+1)$, at the rate $k_2^+$, then the loop length increases to $L+\Delta L_1$ ($\Delta L_1 >0)$. A single extrusion event  ends with ATP hydrolysis (Fig.~\ref{fig:LEmodel}A). We assume that subsequent structural changes  are relatively rapid, returning the chromatin-condensin complex to the starting state with a larger loop. 

It is important to make a distinction between the loop size calculated in our theory and the loop size measured in experiments \cite{ryu2022condensin,ganji2018real}.  The measured extruded loop size is the difference between the total length  and the unextruded length of DNA~\cite{ryu2020condensin}.  As a result, both the states $X_1(i+1)$ and $X_0(i+1)$ (see Figure~\ref{fig:LEmodel}(c)) would have the same loop size in an experiment, even though the newly extruded loop is not fully incorporated into the existing loop. Therefore, when comparing to the experiments, we treat the loop size as $L+\Delta L_i$ in both $X_1(i+1)$ and $X_0(i+1)$, since the length of DNA perceived as ``unextruded" is the same in the two states.  In our model, state $X_0(i+1)$ is poised to take either a forward step, resulting in incorporation of newly extruded loop, or a backward step that effectively decreases the loop size.

During the transition from $X_0(i) \to X_1(i+1)$, condensin captures DNA of length $\Delta L_i$ with $\Delta L_i>0$. Thus, an experiment that  only measures the length of the unextruded DNA segment, will perceive an increase in the loop length by $\Delta L_i$. If the $X_1(i+1)\rightarrow X_0(i)$ transition occurs before the $X_1(i+1)\rightarrow X_0(i+1)$ transition, the loop size ($L$) upon returning to $X_0(i)$ will not increase. However,  an experiment that measures the length of the unextruded DNA will infer the loop size has decreased by $\Delta L_i$(from $L+\Delta L_i$ back to $L$). 

%\st{After a series of cycles of $X_0(i) \to X_1$ (if the $X_1(i+1)\rightarrow X_0(i)$ transition occurs before the $X_1(i+1)\rightarrow X_0(i+1)$ transition, the loop size upon returning to $X_0(i)$ does not increase.) the loop size would increase as $L+\Delta L_1 +\Delta L_2+ ...$ with all $\Delta L_i>0$.  On the other hand, if the $X_1(i+1)\rightarrow X_0(i)$ transition occurs before the $X_1(i+1)\rightarrow X_0(i+1)$ transition, the loop size upon returning to $X_0(i)$ does not increase. In an experiment that  only measures the length of the unextruded DNA segment, the loop size will be perceived to have decreased by $\Delta L_i$ (from $L+\Delta L_i$ back to $L$). This would correspond to a backward step, resulting in loop reversal.}

Conversely, the $X_1(i+1)\rightarrow X_0(i+1)$ transition would render  the loop size unchanged. Thus, in the $i^{th}$ cycle the system first undergoes an apparent forward step of size $\Delta L_i$ upon entering $X_1(i+1)$ from $X_0(i)$. If it then returns to $X_0(i)$ via $X_1(i+1)\!\to\! X_0(i)$, the experiment would register a subsequent backward step of the same magnitude $\Delta L_i$; otherwise, the transition $X_1(i+1)\!\to\! X_0(i+1)$ would be perceived as leaving the loop size unchanged, which would correspond to a futile step in the terminology used in the molecular motor field.

In our picture, the state $X_0(i+1)$ is equivalent to state $X_0(i)$ except that condensin has created a longer loop after undergoing a single catalytic cycle. Moreover, {$X_0(i+1)$} is allosterically equivalent to $X_0(i)$. Thus,  the mapping $X_0(i)\rightarrow$ {$X_0(i+1)$} corresponds to completing one full catalytic cycle while increasing the loop size by $\Delta L_i$.  This is analogous to the walking of conventional motors on F-actin or microtubule, in which upon traversing a specific distance along the polar tracks ($\approx 8$ nm for kinesin\cite{yildiz2004kinesin} or $\approx 36$ nm for myosin V~\cite{Kolomeisky03BJ,Kodera10Nature,Hinczewski13PNAS}) the motor returns to the starting state, ready to begin a new cycle.
%\textcolor{red}{A key difference, however, is that for conventional motors the transition rates within a catalytic cycle are typically insensitive to the absolute position along the track, whereas for condensin the $X_1(i+1)\!\to\! X_0(i+1)$ rate depends on the loop size. Structurally, condensin in $X_1(i+1)$ is the same regardless of loop size, but the loop length modifies the free-energy difference associated with this transition, thereby renormalizing the rate.}

\textbf{Mathematical Model:} We develop a two-state model, based on the physical picture (Fig.~\ref{fig:LEmodel}(c)), which allows us to calculate the distributions of forward ($X_0(i)\rightarrow X_1(i+1)$), reverse ($X_1(i+1)\rightarrow X_0(i)$) steps, and the loop extrusion step ($X_1(i+1)\rightarrow X_0(i+1)$). The rates characterizing these transitions are $k_1^+$ describing the rate for the forward transition, $k_1^-$ describing the rate for the reverse transition, and the rate $k_2^+$ for the irreversible transition {(extrusion step)} $X_1(i+1)\rightarrow X_0(i+1)$. The details of the stochastic kinetic model are given in  Appendix A.

\textbf{Parameters in the theory:}
We denote the number of forward, reverse, and extrusion steps as $m_1$, $m_2$ and $l$ respectively. Magnetic tweezer  experiments  \cite{ryu2022condensin} report the average values of $m_1$ and $m_2$ as a function of an externally applied force. To validate our theory, we first derive the average number of steps $m_1$, $m_2$ or $l$ as a function of the rates (Fig. \ref{fig:LEmodel}).

% We denote the probability of observing the system in either state $I$ or $II$ at time $t$ as $P_I(t)$ or $P_{II}(t)$ respectively. Thus, the master equation for the system described in Fig. \ref{fig:LEmodel} is
% \begin{equation}
%     \frac{dP_I(t)}{dt} = -k_1^+P_I(t)+k_1^-P_{II}(t)+k_2^+P_{II}(t)\text{, and}
% \end{equation}
% \begin{equation}
%     \frac{dP_{II}(t)}{dt} = k_1^+P_I(t)-k_1^-P_{II}(t)-k_2^+P_{II}(t)
% \end{equation}

% Total forward rate of the reaction is given by
% \begin{equation}
%     k^+ = \omega\cdot k_1^+\cdot k_2^+ 
% \end{equation}
% where $\omega$ is a constant that has the dimensions of time. The number of true forward steps during a 2 minute experiment is 
% \begin{equation}
%     m_2 = k^+\cdot(2\text{min})
% \end{equation}
The average number of steps along the  reaction paths in Fig. \ref{fig:LEmodel} during the observation time ($t$) is given in the Table ~\ref{tab:rates}.
\begin{table}[!h]
\begin{center}
\begin{tabular}{| c | c | c |}
\hline
 {Transition} & Rate & Average number of steps \\ 
\hline
\hline 
 $X_0(i)\rightarrow X_1(i+1)$ & $k_{1}^+$ & $m_1=\pi_{X_0}\cdot k_1^+\cdot t$\\
 $X_1(i+1)\rightarrow X_0(i)$ & $k_{1}^-$ & $m_2=\pi_{X_1}\cdot k_1^-\cdot t$\\
 $X_1(i+1)\rightarrow X_0(i+1)$ & $k_{2}^+$ & $l=\pi_{X_1}\cdot k_2^+\cdot t$\\
 \hline
\end{tabular}
\end{center}
\caption{Relating rates in the catalytic cycle to the average number of steps where $t$ is constant. We estimated that in the experiment \cite{ryu2022condensin} $t\approx 50$ s under the condition of $F=0.4$ pN (see Appendix G).}
\label{tab:rates}
\end{table}

In Table \ref{tab:rates}, $\pi_{X_0}$ and $\pi_{X_1}$ are the stationary probability of the system in state $X_0(i)$ and $X_1(i+1)$, respectively. The stochastic kinetic model is used to calculate the rates in our model. 

\textbf{Force-dependent transition rates:}  
We assume that the functional forms of $k_1^+$ and $k_1^-$ are given by the Bell model
~\cite{Bell78Science}, 
\begin{align}
    k_1^+ &= k_{1,0}^+ \exp(-\theta F \Delta R/k_BT) \\ 
    k_1^- &= k_{1,0}^- \exp((1-\theta) F \Delta R/k_BT).
\end{align}

Here, $k_{1,0}^+$, $k_{1,0}^-$ and $\theta$ are adjustable parameters. The average change in hinge-head distance of the condensin molecule in the transition between $X_0(i)$ and $X_1(i+1)$ is $\Delta R=22$ nm, following experimental observations~\cite{ryu2020condensin} and our previous work~\cite{Takaki21NatComm} and $F$ is the external mechanical force. % applied on the motor \textcolor{red}{(since in experiment, the force is applied to DNA rather than motor, this could create some confusion)}.

The transition rate for  $X_1(i+1)\rightarrow X_0(i+1)$ is given by,
\begin{equation}
    k_2^+=k_{2,0}^+\exp(0.58\cdot \Delta L(F) / k_B T),
\end{equation}
\noindent where $k_{2,0}^{+}$ is an adjustable parameter and $\Delta L(F)$ is  force $F$ dependent median length of captured loop. The factor 0.58 is determined by numerically evaluating the change of free energy as a function of the loop size (see Appendix D). It has unit of energy per length which we neglected in the equation.
%\textcolor{red}{(The dimension inside the exponential in Eq.3 is not clear. And should $k_B T$ be in the denominator?)}

\textbf{Dwell time distributions, $P(\tau_R)$ and $P(\tau_F)$:} 
From the experimental observations of LE, it is also possible to calculate the time spent between any two consecutive steps. This is often referred to as the dwell time as it measures the amount of time a molecule ``dwells" in a certain conformation before escaping to a different microstate. Let $\tau_R$ and $\tau_F$  denote the dwell time {before} a backward step and a forward step, respectively. Based on our model, a reverse qstep can only be observed when the system starts at state $X_1(i)$ and undergoes the $X_1(i)\rightarrow X_0(i-1)$ transition. Thus, $\tau_R$ is the time spent in  state  $X_1(i)$ subject to the condition that its next  transition is $X_1(i)\rightarrow X_0(i-1)$. Thus, the dwell time distribution for a backward step is, 
\begin{equation}
    P(\tau_R) = (k_1^-+k_2^+)\exp\left(-(k_1^- + k_2^+)\cdot\tau_R\right).
\end{equation}

The dwell time for a forward step is more nuanced because the observed dwell time before a forward step must include residence in either $X_0(i)$ or the difficult-to-detect intermediate state $X_1(i+1)$, depending on whether the forward step is preceded by a reverse step or another forward step. The full derivation is shown in  Appendix B. The dwell time distribution for forward steps is given by,
\begin{equation}
\label{eq:dwell_time_reverse}
\begin{split}
P(\tau_F) = k_2^+ k_1^+ \frac{e^{-k_1^+ \tau_F}-e^{-(k_1^- + k_2^+)\tau_F}}{k_1^--k_1^+ + k_2^+}+\\
\frac{k_1^- k_1^+}{k_2^++k_1^-}\exp\left(-k_1^+\cdot\tau_F\right).
\end{split}
\end{equation}

\textbf{Determination of the parameters in the model:} The unknown parameters in the theory are $k_{1,0}^+$, $k_{1,0}^-$, $k_{2,0}^+$ and $\theta$. We simultaneously fit the equations, $m_1=\pi_{X_0}\cdot k_1^+\cdot t$, $m_2=\pi_{X_1}\cdot k_1^-\cdot t$ along with the dwell time distributions in Eq. 4 and 5 to experimental measurements \cite{ryu2022condensin} to determine their values. The best fit parameters are given in Table \ref{tab:parameters}.

\begin{table}[!h]
\begin{center}
\begin{tabular}{| c | c |}
\hline
 Parameter & Value \\ 
\hline
\hline 
 $k_{1,0}^+$ & 0.16 s$^{-1}$\\
 $k_{1,0}^-$ &  0.01 s$^{-1}$\\
 $k_{2,0}^+$ &  0.33 s$^{-1}$\\
 $\theta$ & 0.48 \\
 \hline
\end{tabular}
\end{center}
\caption{Best fit parameters of the model. The parameters are obtained by simultaneously fitting the equations, $m_1=\pi_{X_0}\cdot k_1^+\cdot t$ and $m_2=\pi_{X_1}\cdot k_1^-\cdot t$ along with the dwell time distributions in Eq. 4 and 5 to experimental measurements \cite{ryu2022condensin} at $F=0.4$ pN.}
\label{tab:parameters}
\end{table}

\begin{figure*}[ht]
    \centering
    \includegraphics[width=\linewidth]{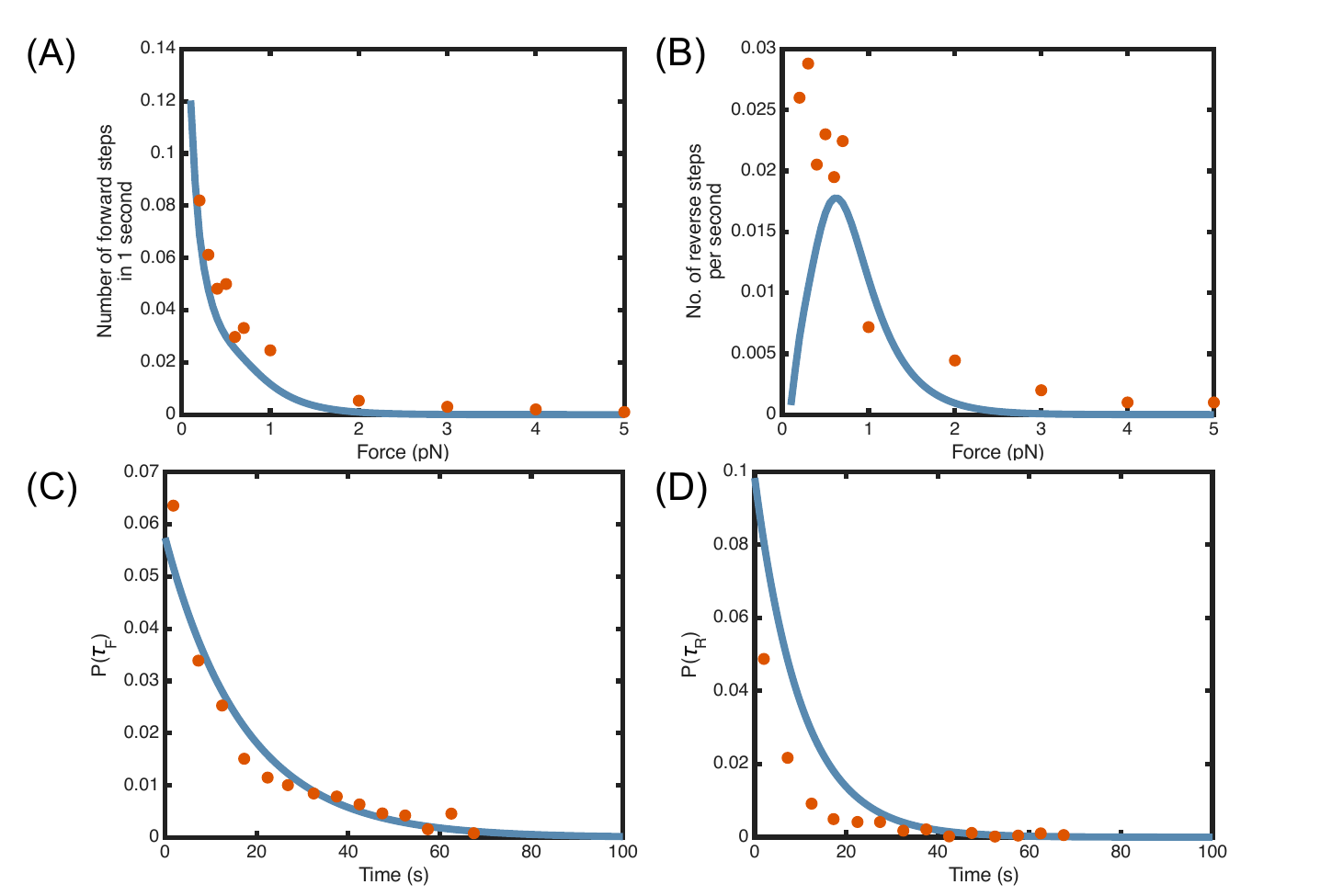}
    \caption{\textbf{Comparison of the stochastic kinetic simulations to experiments:} (A) Number of forward steps traversed in one second ($N_F/s$ a function of force. The blue line is from Gillespie simulations and the dots are from experiments~\cite{ryu2022condensin}. (B) Same as 
    {A} except the results are for the backward steps. Theory predicts a non-monotonic dependence of $N_B/s$ on $F$ at sub-piconewton level forces.  
    (C) Dwell time distribution, $P(\tau_F)$ for the forward steps at $F = 0.4$ pN, calculated using Eq. 4 and 5.    (D) Dwell time distribution $P(\tau_B)$ for the reverse-steps at F$=0.4$ pN. In (C) and (D) the dots are experimental data taken from \cite{ryu2022condensin} reported in Fig. 3F and supplementary Fig. 4B}
    \label{fig:fig2}
\end{figure*}

The fitted and experimental values of the number of forward and reverse steps are shown in Figs. \ref{fig:fig2}(A) and (B) and the  dwell time distribution of the forward and reverse steps are shown in Figs. \ref{fig:fig2}(C) and (D). The fits were stable with respect to random starting conditions for the parameters. The predictions for the dwell time distributions at other values of forces are also plotted (Figs.~\ref{fig:fig3}(C) and (D).  

% \section{\label{sec:level1}Things that don't agree}
% \textbf{Ratio of pseudo reverses and pseudo forward steps:}
% Since we know the average number of pseudo forward steps and pseudo reverse steps taken per unit time, we can compute their ratio using Eq. 4. This result does not match well with the experimental data. 
% \textcolor{red}{Note: For this data the experimental data is highly suspectible to errors during digitization. As we are taking the ratio of two small number, if the image digitization is off by a few pixels the ratio changes substantially. Finally although the last data point at 5 pN is 1, it is not actually so. The point in the graph for the number of pseudo reverse steps at 5 pN was completely covered by the number of forward steps, so I assumed they are equal. Moreover, the experimental error bars for this ratio is quite large, about $0.5$. This is why I asked him to send the data.}
% \begin{figure}[H]
%     \centering
%     \includegraphics[scale=0.6]{RatioOfBackStepAndTransForwardSteps.png}
%     \caption{Ratio of number of backsteps and number of pseudo forward steps as a function of force}
%     \label{fig:my_label}
% \end{figure}
\section{Results}

\textbf{Step-size distribution:}
With the values of the rates in hand, we used the kinetic equations  to compute the properties of the loop extrusion model. We first calculated the step-size distributions of condensin. For conventional molecular motors, like conventional kinesin or myosin V, the step-size distributions are sharply peaked with small dispersions, which imply that these motors take constant steps  as they walk on the polar tracks.  Unlike these motors, forward step-size distribution in condensin is broad~\cite{ryu2022condensin,Strick04CurrBiol,Takaki21NatComm}, which implies that when condensin scrunches DNA to extrude a loop, the length of the extruded segment is highly variable. We previously calculated the broad step-size distributions \cite{Takaki21NatComm} as a function of external load, in quantitative agreement with experiments~\cite{ryu2020condensin,Strick04CurrBiol}. The loop size distribution for the forward steps is broad  because the hinge captures a  segment by making a spatial contact with DNA. Consequently, the hinge-captured segment can be far along the contour from the head-bound site, producing large variability in the extruded contour length. 

Our theory  allows us to calculate the step-size distributions for both forward and reverse steps. To determine the step-size distributions and the velocity of extrusion, we simulated the stochastic kinetic equations for  loop-extrusion based on the Gillespie algorithm~\cite{gillespie1977exact}. The motor transits $X_0(i)$ state at $t=0$, after a time $\Delta t$ where $\Delta t$ is sampled from an exponential distribution with mean $1/k_1^+$. After a successful escape it enters the state $X_1(i+1)$ and captures a loop at the specified external force, defining the (candidate) step size. The motor exits state $X_1(i+1)$, after $\Delta t$ that is sampled from another exponential distribution with mean $1/(k_1^-+k_2^+)$. Upon conditional on exiting, it returns to state $X_0(i)$ with probability $k_1^-/(k_1^-+k_2^+)$, which we record as a reverse step, a negative step size,  (the captured loop is not incorporated. Alternatively, the motor  proceeds to $X_0(i+1)$, with probability $k_2^+/(k_1^-+k_2^+)$, which we record as a forward step (the captured loop is incorporated). Subsequently, the chemical cycle is reset and the process repeats from state $X_0(i+1)$. %\textcolor{purple}{(I am still confused about the algorithm. In Gillespie algorithm, $\Delta t$ is sampled from exponential distribution whose mean is inverse of total rates. Here, the mean is fixed to be 1 second. If this is case, this is still not technically Gilliespie.)}

\begin{figure*}[ht]
    \centering
    \includegraphics[width=\linewidth,trim={0 200 0 100}]{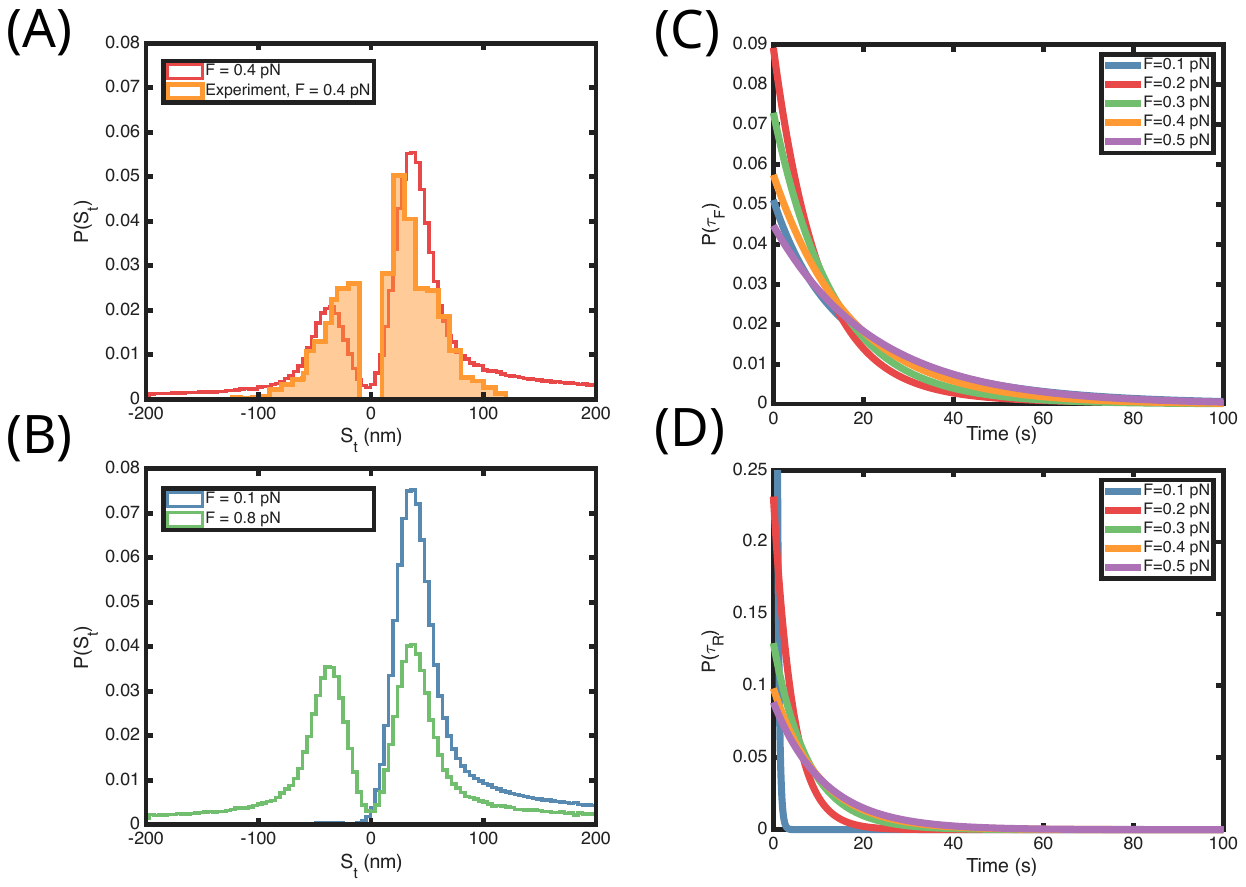}
    \caption{\textbf{Step-size distributions and dwell time distributions:}  (A) The solid line is theoretical prediction  and the shaded histogram is taken from experiments~\cite{ryu2022condensin} at  $F=0.4$ pN.  (B) Predicted step size distributions at  $F=0.1$ pN and $F=0.8$ pN. For both step-size distribution calculations, unlike the convention used by Ryu \textit{et al.}~\cite{ryu2022condensin} we plot the forward steps on the positive X axis and reverse steps on the negative X axis. This is done to keep the notation consistent with the  literature convention on molecular motors.
    (C) Dwell time distributions  for forward steps as a function of $F$ calculated using the stochastic kinetic model.  (D) Same as except (C) the results are for backward steps. The distributions vary monotonically as a function of $F$ unlike the results in (C).
    }
    \label{fig:fig3}
\end{figure*}

The step-size distribution at different values of force  along with a corresponding experimental measurements are shown in (Fig.~\ref{fig:fig3}). At the external force of $F=$0.4 pN, the theoretical predictions is in good agreement with the experimental measurements, as shown in Fig.~\ref{fig:fig3}.  Moreover, as $F$ increases, the reverse steps become more prominent and the  step-size distribution progressively approaches that for forward steps, resulting in an increasingly symmetry between the forward and reverse distributions (Fig.\ref{fig:fig3}(B)). This differs from kinesin and myosin, where increasing load makes the reverse steps sufficiently frequent. As a result,  the reverse-step distribution eventually dominates and skews the overall stepping statistics toward backward motion, as predicted for kinesin~\cite{vu2016discrete}. %(I hope I understand the original text correctly. I rewrote this to make the point clear. Also I think to make this point more convincble, we need to show when force increases, the reverse distribution does not eventually dominate the forward one.)}

\textbf{Catch bond characteristics in condensin:}
Several protein complexes exhibit catch bond behavior presence of force~\cite{dembo1988reaction,Marshall03Nature,Barsegov05PNAS,Choi25AnnRevImm,Thomas2008ARB}. It is generally expected that application of an external force should lower the free energy barrier to bond-rupture in accord with the Bell model~\cite{Bell78Science}, thus decreasing its lifetime.  However, it was hypothesized sometime ago~\cite{dembo1988reaction,dembo1994lectures} that external forces could increase the lifetime of a bond by stabilizing an alternative conformation. These two different responses of complexes in the presence of an external force are referred to slip (lifetime decreases with force) and catch bonds (lifetime increases with force), respectively. In a catch bond force application leads to an increase in the lifetime of an intermediate state up to a critical value of force~\cite{Barsegov05PNAS,chakrabarti2017phenomenological,Barsegov06JPCB} beyond the lifetime decreases as force continues to increase. This counter intuitive non-monotonic response of the lifetime of the complex to external force is the distinguishing feature of a catch bond. We show below that our theory predicts  that  condensin motor also exhibits a similar non-monotonic response to external force, leading to the conclusion that state $X_1$ is in a catch bond state.

First, let us highlight the unique role played by state $X_1$. There are two ways to exit state $X_1$: $X_1(i+1)\rightarrow X_0(i)$ transition with rate $k_1^-$ (reverse step) and the $X_1(i+1)\rightarrow X_0(i+1)$ transition with rate $k_2^+$ (extrusion step). The rates $k_1^-$  and $k_2^+$ exhibit contrasting behavior as a function of $F$. As indicated by the positive value of $\theta$, the rate $k_1^-$ increases as force increases, key feature of the catch bond.  Intuitively, this means that the scrunching process becomes more difficult at higher forces and is prone to losing the extruded chromatin loop length by the motor taking a backward step.   On the other hand, the rate $k_2^+$ is a function of the length of the new loop that is incorporated into the existing loop. At high external forces the length of the loop captured is shorter, hence the free energy advantage of increasing the size of the loop is reduced. The rate $k_2^+$ thus decreases at high $F$ (slip bond) until the motor stalls.
Thus, among the two transitions $X_1(i+1)\rightarrow X_0(i)$ and $X_1(i+1)\rightarrow X_0(i+1)$, the rate of $X_1(i+1)\rightarrow X_0(i)$ increases with increasing external force, while the rate of $X_1(i+1)\rightarrow X_0(i+1)$ decreases with increasing external force. This opposing force response of the two rates creates a situation where increasing the external force up to a critical value $\approx$ 0.4 pN increases the lifetime of state $X_1(i+1)$, creating a catch bond. Upon further increase of the external force the lifetime of $X_1(i+1)$ decreases again, which is the characteristic of a slip bond. 

\begin{figure*}[ht]
    \centering    \includegraphics[width=\linewidth,trim={0 0 0 0}]{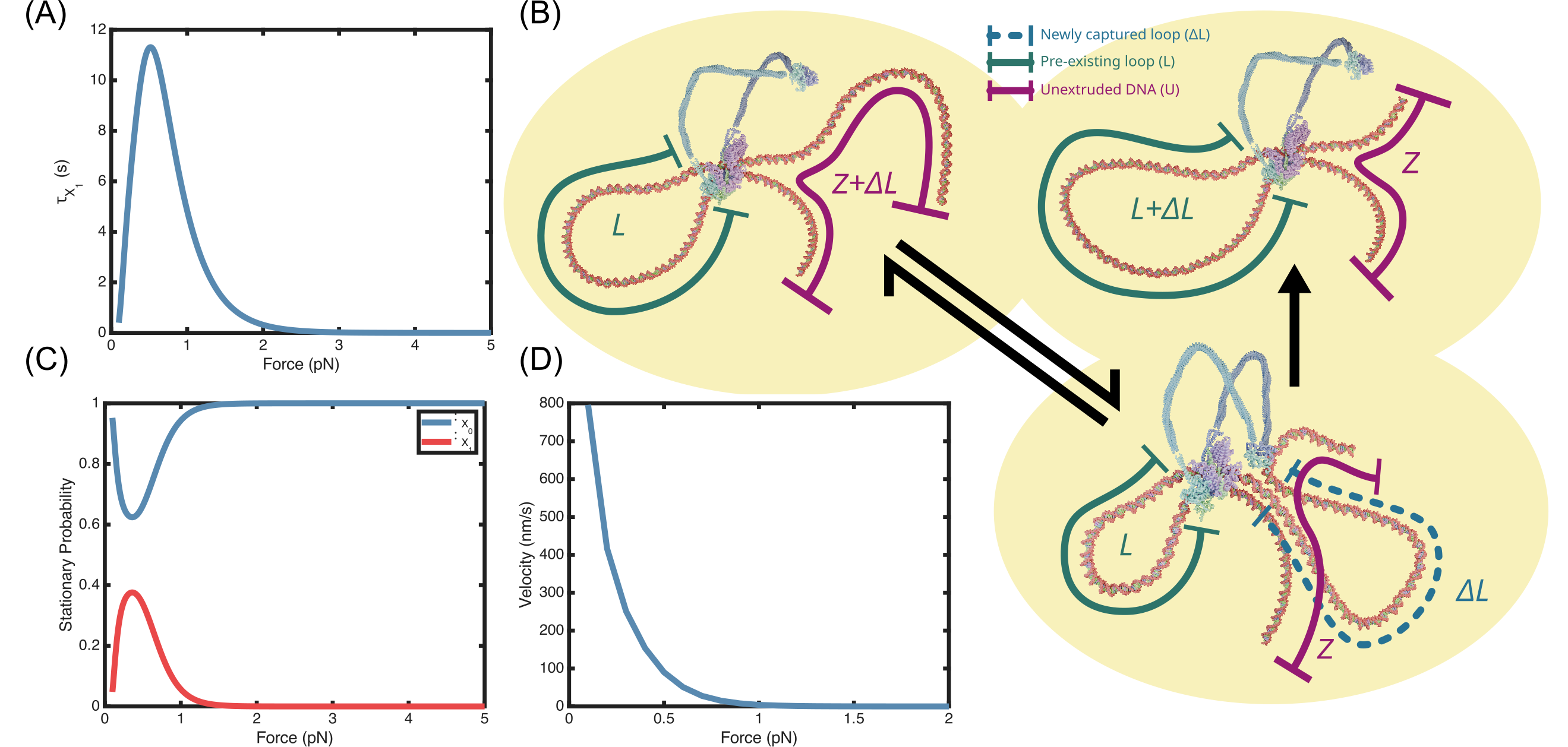}
    \caption{\textbf{Catch Bonds:} (A) Lifetime of state $X_1$ that is poised to take a forward or backward step as a function of force, $F$. The increase in $\tau_{X_1}$ at low $F$ corresponds to catch bonds and the subsequent decrease at higher $F$ results from the slip bond behavior. (B) Rendition of states $X_0 (i)$ and $X_0 (i+1)$ associated with loop extrusion. Structure for the head and hinge domain were prepared based on PDB:7QEN and PDB: 6YVU, respectively. DNA and the coiled-coil domains are schematics. In the $X_1(i+1)$ state the length of the unextruded chromatin reduces by $\Delta L$. (C) Stationary probability of states $X_0$ and $X_1$ as a function of  $F$.  (D) Predicted force-velocity curve of the loop extrusion using the theory proposed here. The zero-force value ($\sim$ 300 nm/s) and the stall force ($\sim$ 0.8 pN) are in very good agreement with experiments.}
\label{fig:IIlifetime}
\end{figure*}

The catch bond feature of state $X_1(i+1)$  follows from the dependence of its lifetime as a function of the external force. The average lifetime of state $X_1(i+1)$ is,

\begin{equation}
    \tau_{X_1(i+1)} = 1/(k_1^-+k_2^+),
\end{equation} 

\noindent where $k_1^-+k_2^+$ is the total transition rate from $X_1(i+1)$ state. Using the parameters in Table \ref{tab:parameters}, $\tau_{X_1(i+1)}$ in Fig. \ref{fig:IIlifetime}(A) shows an increase till $F\approx$ 0.4 pN, which is roughly half the value of the stall force~\cite{ganji2018real}.  The calculated  stationary probability $\pi_{X_1(i+1)}$ in Fig. \ref{fig:IIlifetime}(C)  shows that although the motor predominantly stays in state $X_0(i)$, the stationary probability of being in state $X_1(i+1)$ also changes non-monotonically as a function of force.

\textbf{Experimental validation:}
Ultimately  the theoretical predictions, especially those pertaining to the force-dependent kinetics and the emergence of catch-bond behavior,  should be validated by future experiments. We delineate a strategy for testing the theoretical predictions. In particular, a central prediction of the model is the dependence of the microscopic rates $k_1^+(F)$, $k_1^-(F)$, and $k_2^+(F)$ on the applied force $F$. Because these rates can be extracted from experiments at various forces they can be compared with our theory.

As previously mentioned,  the dwell time of the reverse step is exponentially distributed with the mean rate, $k_1^-$. Thus, the average dwell time for a reverse step as a function of force, is $1/k_1^-$. Similarly, the conditional dwell-time distribution of a forward step immediately after a reverse step has an exponential distribution with rate $k_1^+$. 
As shown in Appendix F, the average conditional dwell time for a forward step immediately after another forward step is $(k_1^-+k_1^++k_2^+)/(k_1^+(k_1^-+k_2^+))$. These three measurements are sufficient to determine all three rate constants in the model, which is could be used to verify our predictions.

\section{Discussion}
We developed a theory to explain the results of single molecule magnetic tweezer experiments performed at a fixed ATP concentration, demonstrating  that condensin extrudes loops both in the forward and backward directions~\cite{ryu2022condensin}. Using a stochastic kinetic model and the scrunching mechanism~\cite{Revyakin06Science,Kapanidis06Science} as the basis of loop     extrusion~\cite{ryu2020condensin,Takaki21NatComm},  we quantitatively accounted for the key experimental observations~\cite{ryu2022condensin}. Let us first summarize the major results: (1) The theory has four unknown parameters, which were determined data at a single value of force. The theory is used to predict the measured dwell time distribution for forward and reverse steps at other values of force. (2) Without any further adjustments to the rate constants obtained at a single force value,  the theory also accounts for the step-size distributions ($P(S_t)$s)  in both the forward and backward steps. The calculated $P(S_t)$s  is in excellent agreement with experiments~\cite{ryu2022condensin}.  Predictions for $P(S_t)$s at other values of $F$ also follow from the theory.   (3)  The dwell time distribution for the forward step, $P(\tau_F)$ as a function of $F$ differs \textit{qualitatively} from $P(\tau_R)$ for the backward step. The former changes non-monotonically as $F$ increases whereas  $P(\tau_R)$ decreases monotonically  at all force values.  Strikingly, there is a maximum in $P(\tau_F)$ at short $\tau_F$, suggestive of  catch bond behavior, which is the major prediction of the theory. The lifetime of the state that has the propensity to take either step forward or backward increases at low forces (catch bond) and decreases (slip bond) at higher forces.

\textbf{Non-Markovian nature of backward steps: } In contrast to conventional motors, condensin  takes backwards steps \textbf{only} immediately after one or more preceding forward steps~\cite{ryu2020condensin}. Loop reversal is reflected in the non-exponential nature of the dwell time distribution (Eq.\ref{eq:dwell_time_reverse}). The condition for loop reversal that is predicated on the forward steps suggests that condensin has short term ``memory" of the steps. Notably, Ryu \textit{et al.}~\cite{ryu2022condensin} report a strong correlation between the size of a backward step and the one immediately preceding the forward step (Fig.~3D in ~\cite{ryu2022condensin}), supporting this interpretation. If this key finding is established with additional experiments, it would imply  that a backward step negates, possibly only partially, the loop length gain of a only the single preceding forward step.   It would have no effect on the length of the loop extruded  by all other forward steps  before the backward step.  

\textbf{Condensin takes long steps in both directions:} It is noteworthy that condensin extrudes loops whose lengths frequently exceed  the size of the SMC complexes. The heads and the hinge, which are connected by semi-flexible coiled coils, are separated by $\sim$ 50 nm in the \textit{apo} state.  Upon ATP binding the head-hinge distance decreases by an average of $\approx$ 22 nm~\cite{ryu2020condensin} through an allosteric scrunching mechanism. However, the size of the extruded loops could exceed 100 nm~\cite{Strick04CurrBiol,Takaki21NatComm,ryu2022condensin}. 
There are very few high-resolution structures of the SMC complexes, making the structural basis for the large sizes of the extruded loops unclear.  By comparing with conventional motors, which move on stiff tracks (microtubules or F-actin), the DNA ``track" is highly flexible. We suggest vast differences in the track stiffness between DNA (persistence length $\approx$ 50 nm) compared to F-actin ($l_p \approx 1\ \mu \text{m}$) or microtubule ($l_p  \approx$ 1 mm) may be reason for the variability  and the large step sizes taken by the condensin both in the forward and the reverse steps.        
%Moreover, as condensin is a molecular motor its structures can exist in multiple conformational states under similar experimental conditions. Most condensin structures have been observed only in the ATP bound state. This dearth of experimental structures constrained our modeling approach to a kinetic model.

\textbf{On the utility  of kinetic models:} Kinetic models have been successfully used to explain the stepping dynamics of conventional motors~\cite{Kolomeisky03BJ,kolomeisky2007molecular,Mugnai20RMP} as well as catch bonds in cell adhesion and actomyosin complexes~\cite{Marshall03Nature,Barsegov05PNAS,Barsegov06JPCB,chakrabarti2017phenomenological}. For instance, theories based on the stochastic kinetic models have been used to calculate the distribution of force-induced unbinding lifetimes of cell adhesion complexes~\cite{Barsegov05PNAS}. Furthermore, the model also predicted, in accord with experiments~\cite{Harder15BJ}, that generically, in such systems one ought to find slip bonds at very low forces followed by catch bond at intermediate $F$ and slip bonds (slip $\rightarrow$ catch $\rightarrow $ slip) at high $F$~\cite{Harder15BJ}. Despite these successes a natural criticism of the stochastic kinetic models is that they could  contain several rate constants.  For each kinetic step, there are two kinetic rates accounting for the forward and backward reactions. As the number of steps in a reaction scheme increases the parameters needed needed to fit the experimental data increase. This understandable criticism should be weighed against the successful use of kinetic models in biophysics and more generally in chemistry. 
By building on this rich history,  the proposed a minimal  model of loop extrusion by condensin simultaneously explains experiments for both forward and backward steps.  An important feature of our minimal model is that it uses only four adjustable parameters to describe the loop extrusion process. The parameters are uniquely determined fitting to the  model to the experiments,  namely the number of forward and backward steps and their respective dwell time distributions. A striking experimentally verifiable prediction is that the backward step involves an intermediate state that has catch bond characteristics. The simplicity of the kinetic model also allows us to propose schematics of intermediates of condensin driven loop extrusion, as shown in Fig. 4 (B). Such structures can be further refined as additional experimental data (cryo-EM for instance) become available.

\textbf{Catch Bonds:}  There are several passive systems, such as cell-adhesion complexes, where the prevalence of catch bonds has been documented. However, there is no active system in which the existence of catch bonds are been been demonstrated. The exception is the study (largely theoretical) involving  cytoplasmic dynein where the increase in the motor velocity at low ATP concentration in the force range of (1-3) pN~\cite{Johnson20PhyBiol} has been interpreted as evidence of catch bonds. However, a direct measurement of the force-dependent lifetime of the state in dynein or other active system is lacking possibly due to the absence of structures accessed in the ATPase cycles. This is also the case in SMC motors. On the other hand,  theories based on the structures of cell adhesion complexes, in the presence and absence of force, show that if mechanical force  leads to   an alternate stable state, it would result in the catch bond formation~\cite{Barsegov05PNAS,Chakrabarti14PNAS}. Such is the case in the P-Selectin bound to the glycoprotein PSGL-1, which exists in an bent state in the absence of force and forms an extended state in the presence of force. With the benefit of knowing the crystal structures in the bent and extended states, a microscopic  theory has been formulated~\cite{Chakrabarti14PNAS,Barkan24PNAS} to explain catch bond formation in this passive system.  This has not been possible in motors. 

The emergence of catch bonds predicted in this work hinges on the assumption that a state that is poised to take either forward or backward step exists (Fig. \ref{fig:LEmodel}) in the catalytic cycle.  Although not detected in experiments, the existence of such a state might also suggest that the motor should occasionally pause which could possibly have biological significance~\cite{Dey23CellReports}. 

%The tendency to find catch bonds Presence of a catch bond is generally associated with resistance to external forces. Nevertheless, a natural criticism of stochastic kinetic models that could contain several rate constants, which is sometimes viewed as a limitation.  We believe a catch bond like interaction may be present in condensin molecules to provide them some resilience to external force. In the presence of increased external force, the lifetime of state $\text{II}$ goes up, implying that condensin is paused in the middle of its extrusion cycle.

%\textcolor{blue}{The results of our study implies that reverse stepping of condensin goes through a paused state which could possibly have biological significance.
%6. Why does our calculation suggest that the mechanism of reversal is a catch bond?
%8. Total timescale for full reaction cycle. Maybe compare with extrusion velocity?}
\section{Conclusion}

We introduced a stochastic kinetic model for loop extrusion by condensin that accurately reproduces the experimental behavior for both forward and reverse steps. The model builds on the assumption that condensin fluctuates between two conformational states while extruding loops by the scrunching mechanism.   The theory shows that condensin should exhibit catch bond-like behavior (prominent during loop reversal) where upon application of an external force the lifetime of the intermediate states increases. The novel prediction is amenable to experimental test.

% \begin{figure}[ht]
%     \centering
%     \includegraphics[width=\linewidth,trim={0 0 0 0}]{Figures/MedianDeltaL.pdf}
%     \caption{Median $\Delta L$ as a function of external force. The inset shows how $P(\Delta L)$ changes with external force. }
%     \label{fig:MeanCapLength}
% \end{figure}

% \begin{figure}[ht]
%     \centering
%     \includegraphics[width=0.9\linewidth,trim={100 170 100 100}]{Figures/DwellTimeOtherForce.pdf}
%     \caption{Dwell time distributions of (A) forward-steps and (B) reverse-steps, at other values of force.}
%     \label{fig:my_label}
% \end{figure}

% \begin{figure}[ht]
%     \centering    \includegraphics[width=\linewidth,trim={0 250 0 100}]{Figures/CondensinLoopStructures.pdf}
%     \caption{Rendition of two different microstates of loop extrusion (A) state $X_0 (i)$ and $X_0 (i+1)$ and (B) state $X_1 (i+1)$ and after it slipped back to $X_0 (i)$. The structure for the head and hinge domain were prepared using PDB:7QEN and PDB: 6YVU respectively. The DNA and the coiled-coil domains are not based on real structures. In state $X_1(i+1)$ the length of the unextruded part of the DNA reduces by $\Delta L$}
%     \label{fig:Structure}
% \end{figure}
\nocite{*}
\FloatBarrier
\section{Appendix}
\subsection{Stochastic Kinetic Model}
The catalytic cycle (Figure \ref{fig:LEmodel}) shows that after the state $X_0(i+1)$ is reached, the motor is back to the initial state, and a new cycle begins. By noting the equivalence of the states $X_0(i)$ and $X_0(i+1)$ to which ATP is bound, the time dependent changes in the populations, $P_{X_0}(t)$ and $P_{X_1}(t)$, are given by the kinetic equations,
\begin{align}
\frac{dP_{X_0}(t)}{dt} &= -k_1^+P_{X_0}(t)+k_1^-P_{X_1}(t)+k_2^+P_{X_1}(t)\\
\frac{dP_{X_1}(t)}{dt} &= k_1^+P_{X_0}(t)-k_1^-P_{X_1}(t)-k_2^+P_{X_1}(t).
\end{align}

%\begin{equation}
%    \frac{dP_{X_0}(t)}{dt} = -k_1^+P_{X_0}(t)+k_1^-P_{X_1}(t)+k_2^+P_{X_1}(t)\text{, and}
%\end{equation}

%\begin{equation}
%    \frac{dP_{X_1}(t)}{dt} = k_1^+P_{X_0}(t)-k_1^-P_{X_1}(t)-k_2^+P_{X_1}(t).
%\end{equation}

\noindent The steady state condition, $\text{d}P_{X_0}(t) / \text{d}t=\text{d}P_{X_1}(t) / \text{d}t=0$, leads to,
\begin{equation}
    k_1^+\pi_{X_0}-(k_1^-+k_2^+)\pi_{X_1}=0,
\end{equation}

\noindent where $\pi_{X_0}$ and $\pi_{X_1}$ are the stationary probabilities of observing the states $X_0$ and $X_1$, respectively. Using the condition $\pi_{X_0}+\pi_{X_1}=1$, the stationary probabilities for observing the system in either state $X_0$ or $X_1$ are,
\begin{align}
    \pi_{X_0} &= \frac{k_2^++k_1^-}{k_2^++k_1^-+k_1^+} \\
    \pi_{X_1} &= \frac{k_1^+}{k_2^++k_1^-+k_1^+}.
\end{align}

\subsection{Forward step dwell time distribution}
The kinetic scheme shows that a forward step  occurs when the system undergoes the $X_0(i)\rightarrow X_1(i+1)$ transition. The transition $X_1(i+1)\rightarrow X_0(i+1)$ cannot be directly detected in single molecule experiments, as it does not result in a change in the length of the unextruded DNA. As a result,  before taking a forward-step, condensin could be either in  $X_0(i)$ or in  $X_1(i+1)$. These two cases can be distinguished as follows. (a) If the current forward-step is preceded by a reverse-step then the system would have been only in state $X_0(i)$ and the forward step happens via the $X_0(i)\rightarrow X_1(i+1)$ transition. The dwell time distribution for these forward steps is $k_1^+\exp\left(-k_1^+\cdot\tau_F\right)$. (b) However, if the forward step, is preceded by another forward step, then according to our model, the system started in state $X_1(i)$, then moved to state $X_0(i+1)$ through the transitions $X_1(i)\rightarrow X_0(i+1)$, and which would be followed by another transition $X_0(i+1)\rightarrow X_1(i+2)$. Thus, the time to take a forward-step after another forward-step is a hypoexponential (also known as generalized Erlang) distribution. The exact form  is derived in the Appendix F. The total dwell time distribution is a combination of these two possibilities, weighted by the probability of observing a forward-step or reverse-step. The dwell time distribution for forward-steps is given by,
\begin{equation}
\begin{split}
P(\tau_F) = k_2^+ k_1^+ \frac{e^{-k_1^+ \tau_F}-e^{-(k_1^- + k_2^+)\tau_F}}{k_1^--k_1^+ + k_2^+}+\\
\frac{k_1^-k_1^+}{k_2^++k_1^-}\exp\left(-k_1^+\cdot\tau_F\right).
\end{split}
\end{equation}

\subsection{Free energy change for loop incorporation in a Rouse polymer}
Our goal is to obtain the free-energy change associated with loop incorporation in a Rouse polymer \cite{doi1988theory}. Specifically, we compute the free-energy difference between states $X_0$ and $X_1$, $\Delta G_{X_0(i)|X_1(i+1)}$, by evaluating the partition function of a ring Rouse chain with bond length $b$ and $N$ bonds. The model Hamiltonian is,
\begin{equation}
    \mathcal{H} = \frac{3k_BT}{2b^2}\left[(\textbf{r}_{N}-\textbf{r}_1)^2+\sum_{i=1}^{N-1} (\textbf{r}_{i+1}-\textbf{r}_i)^2\right].
\end{equation}

To preserve  translational invariance, we modified the Hamiltonian by tethering the first monomer to the origin giving us,
\begin{equation}
    \mathcal{H} = \frac{3k_BT}{2b^2}\left[\textbf{r}_1^2+(\textbf{r}_{N}-\textbf{r}_1)^2+\sum_{i=1}^{N-1} (\textbf{r}_{i+1}-\textbf{r}_i)^2\right].
\end{equation}
The partition function, 
\begin{equation}
    \Xi = \int \exp\left[-\mathcal{H}/k_BT\right],
\end{equation}
can be written as,  $\Xi = \Xi_x^3$, where $\Xi_x\sim\int \exp\left[-\mathcal{H}_x/k_BT\right]$; $\mathcal{H}_x$ is the x-component of the separable Hamiltonian, computed using the x-coordinates of the monomers. $\Xi$ can be written as a multivariate Gaussian integral as $\Xi = \int \exp\left[-(1/2)x^TAx\right]$, where $A$ is a matrix,
\begin{equation}
\textbf{A}_{ij}= 
\begin{cases}
    3\left(\frac{3}{b^2}\right),      & \text{if } i=j=1 \\
    2\left(\frac{3}{b^2}\right),      & \text{if } i=j\neq1 \\
    -1\left(\frac{3}{b^2}\right),     & \text{if  $|i-j|=1$}\\
    -1\left(\frac{3}{b^2}\right),     & \text{if  $i=1,j=N$ or $i=N,j=1$}\\
    0,      & \text{otherwise}.
\end{cases}
\label{eq:connectivity}
\end{equation}
Evaluation of the integral yields, $\Xi_x=(2\pi)^{N/2}|A|^{-1/2}$. This simplifies to $\Xi_x=(2\pi)^{N/2}(3/b^2)^{-N/2}N^{-1/2}$. Then we can write 
\begin{equation}
    \Xi = \left(\frac{2\pi b^2}{3}\right)^{3N/2}N^{-3/2}
\end{equation}
The free energy of a ring Rouse polymer with $N$ monomers, modeled as Rouse polymer is, 
\begin{equation}
    G = \frac{3k_BT}{2}\ln\left(N\right)-\frac{3Nk_BT}{2}\ln\left(\frac{2\pi b^2}{3}\right)
\end{equation}
Thus, the free energy change of incorporating a new segment of length $\Delta L$ is given by,
\begin{equation}
    \Delta G_{X_1(i+1)|X_0(i)} = \frac{3}{2}k_BT\left[ \ln\left(1+\frac{\Delta L}{N_L\cdot b}\right)-\frac{\Delta L}{b}\ln\left(\frac{2\pi b^2}{3}\right)\right]
\end{equation} 
In the limit of a large pre-existing loop, $N_L \gg \Delta L/b$, the first term is negligible. The resulting free-energy change upon incorporating an additional loop varies approximately linearly with the incorporated loop size, $\Delta L$.

\subsection{Free energy change for loop incorporation of semi-flexible polymers}

For a semi-flexible polymer it is not possible to derive an  exact expression for the free energy as a function of polymer length, so we rely on some approximations. We assume the loop is a semi-flexible polymer ring of length $L$ with bond-length $a$ and persistence length $l_p$. The number of segments in this ring is $N=L/a$. Following an earlier work~\cite{ha1995mean}, we write the Hamiltonian for the ring semi-flexible polymer as,
\begin{equation}
\begin{split}
        \mathcal{H} = \frac{1}{2}\left[\frac{l_p}{a^3}\sum_{i=1}^{N} (\Delta\textbf{r}_{i+1}-\Delta\textbf{r}_i)^2+\frac{2\lambda}{a}\Delta\textbf{r}_i\right]-\lambda\cdot a\cdot N,
        \label{eq:semHamiltonian}
\end{split}
\end{equation}
where $\Delta\textbf{r}_i=\textbf{r}_{i+1}-\textbf{r}_i$ and the index $N+1=1$ to ensure ring closure. The optimal value of $\lambda$, which reproduces the statistical properties of semi-flexible chains, is calculated using mean field theory. The resulting expression is given by~\cite{ha1995mean},
\begin{equation}
    \left(\frac{l_p\lambda}{2}\right)^{1/2}=\frac{3}{4}\left[\coth\left\{ L\left(\frac{\lambda}{2l_p}\right)^{1/2}\right\} -\frac{1}{L}\left(\frac{2l_p}{\lambda}\right)^{1/2}\right]
    \label{eq:lambda}    
\end{equation}

To obtain the free energy of a semiflexible polymer as a function of its length, we numerically calculate $\lambda$ using Eq. \ref{eq:lambda} and substitute that back into Eq. \ref{eq:semHamiltonian}. 

The mean field free energy is  $G_{\text{MF}}=-\ln \left[\int\Pi_n d\textbf{r}_n \exp(-\mathcal{H}[\{\textbf{r}_n\}])\right]$ follows from and is given by,
\begin{equation}
    G_{\text{MF}} = \frac{3}{2}\ln(|\textbf{Q}|)-\frac{3N}{2}\ln\left[\frac{2\pi}{lp/a^3+2\lambda/a}\right]-\lambda\cdot a\cdot N.
\end{equation}
In the above equation, $\textbf{Q}=(\textbf{A}+\textbf{B})/(lp/a^3+2\lambda/a)$ is the sum of the two matrices containing the angle and bond information, which we denote by \textbf{A} and \textbf{B} respectively. 
The matrix \textbf{A} can be written as,
\begin{equation}
\textbf{A}_{ij}= 
\begin{cases}
    6\left(\frac{l_p}{a^3}\right),      & \text{if } i=j \\
    -4\left(\frac{l_p}{a^3}\right),     & \text{if  $|i-j|=1$}\\
    1\left(\frac{l_p}{a^3}\right),     & \text{if  $|i-j|=2$}\\
    -4\left(\frac{l_p}{a^3}\right),     & \text{if } \{i,j\}=\{1,N\} \text{ or } \{N,1\}\\
    1\left(\frac{l_p}{a^3}\right),     & \text{if } \{i,j\}=\{1,N-1\}\text{,}\{2,N\}\text{,}\\
    &\{N-1,1\}\text{ or }\{N,2\}\\
    0,      & \text{otherwise}
\end{cases}
\label{eq:semAngles}
\end{equation}
Similarly, \textbf{B} can be written as
\begin{equation}
\textbf{B}_{ij}= 
\begin{cases}
    2\left(\frac{2\lambda}{a}\right),      & \text{if } i=j \\
    -1\left(\frac{2\lambda}{a}\right),     & \text{if  $|i-j|=1$}\\
    -1\left(\frac{2\lambda}{a}\right),     & \text{if } \{i,j\}=\{1,N\} \text{or} \{N,1\}\\
    0,      & \text{otherwise}.
\end{cases}
\label{eq:semBonds}
\end{equation}
Using these relations, we plot the mean field free energy of a semi-flexible ring as a function of contour length. Using $a=3.4$ nm (the length of 10 base pairs of DNA) and $l_p=50$ nm (nominal persistence length of dsDNA), we plot the mean field free energy as a function of the length of the loop.
\begin{figure}[H]
    \centering
    \includegraphics[scale=0.6]{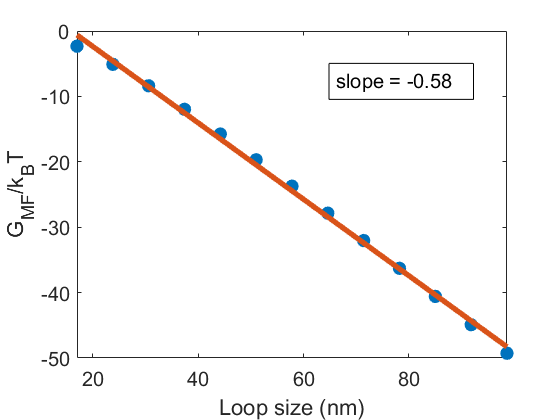}
    \caption{Mean field free energy as a function of the size of loop. Slope is obtained by fitting to the numerical results.}
    \label{fig:semiFreeEnergy}
\end{figure}
The linear dependence of the free energy on the size of the loop, as shown in Fig.~\ref{fig:semiFreeEnergy} implies that when the loop size changes by $\Delta L$, the free energy changes as $\Delta G_{\text{MF}}\approx0.58\cdot \Delta L$. 

\subsection{Parametrization of the rate constants}
The equilibrium constant $K_{\text{eq}}$ for the $X_0(i) \rightleftharpoons X_1(i+1)$ process is given by $K_{\text{eq}} = k_1^+/k_1^- = \exp(-\Delta G_{X_1(i+1)|X_0(i)}F/k_BT)$, where $\Delta G_{X_1(i+1)|X_0(i)}(F)$ is the free energy difference between states $X_0(i)$ and $X_1(i+1)$ in the presence of an external force $F$. We  rewrite the free energy as $\Delta G_{X_1(i+1)|X_0(i)}(F)=\Delta G_{X_1(i+1)|X_0(i)}(0)+F\Delta R$, where $F\Delta R$ is the work done by condensin to change its size by $\Delta R$, and $\Delta G_{X_1(i+1)|X_0(i)}(0)$ is the free-energy difference between state $X_0(i)$ and $X_1(i+1)$ of the condensin DNA complex in the 0 force. We write $\exp(-\Delta G_{X_1(i+1)|X_0(i)}(0)/k_BT)=k_{1,0}^+/k_{1,0}^-$ where $k_{1,0}^+$ and $k_{1,0}^-$ are the rates of the transitions $X_0(i)\rightarrow X_1(i+1)$ and $X_1(i+1)\rightarrow X_0(i)$  respectively, at zero force $F=0$. 

\subsection{Dwell time for the forward step}

We model the reaction path $X_1(i+1)\rightarrow X_0(i+1)$ as an irreversible process, allowing us to approximate the rate $k_2^+$ as:
\begin{equation}
    k_2^+ \sim\exp(-\Delta G_{X_0(i+1)|X_1(i+1)}(F)/{k_BT}),
\end{equation}
where $\Delta G_{X_0(i+1)|X_1(i+1)}(F)$ is the free energy cost for loop incorporation in the presence of an external force $F$. We estimate $\Delta G_{X_0(i+1)|X_1(i+1)}(F)$ as the free energy difference change the contour length a semi-flexible ring polymer increases  from $L$ to $L+\Delta L(F)$, where $\Delta L(F)$ is the length of the new loop captured or the step-size of condensin in the presence of force $F$. 

We previously derived the distribution for $\Delta L(F)$ in \cite{Takaki21NatComm}.  We  show in the Appendices C and D that the free-energy of both flexible and semi-flexible ring polymers increase linearly with the length of the polymer. Assuming that the size of each monomer is $a=3.4$ nm (the length of 10 base pairs of DNA) and the persistence length of the polymer $l_p=50$ nm (persistence length of dsDNA) the free-energy of the ring polymer decreases as $\Delta G_{X_0(i+1)|X_1(i+1)}\sim -0.58\cdot\Delta L(F)\cdot k_BT$. Thus the rate of transition along the  $X_1(i+1)\rightarrow X_0(i+1)$ reaction can be written as:
\begin{equation}
    k_2^+=k_{2,0}^+\exp\big(C_1\cdot \Delta L(F)/k_B T\big),
\end{equation}
where $C_1 = 0.58$. The above equation tends generate large values, especially when $\Delta L(F)$ big,  which makes the numerical evaluation of the exponential unstable. To avoid this, we rescaled the equation above as $k_2^+=k_{2,0}^+\exp(0.58\cdot (\Delta L(F)-52)\cdot k_BT)$, where $52$ nm is approximately the value of median $\Delta L$ at 0 force (see Fig. 2(a) in~\cite{Takaki21NatComm}.) Thus, in the rescaled equation $k_{2,0}^+$ is the rate at 0 force which we treat as an adjustable parameter. In principle, $\Delta L(F)$ can be sampled from the step-size distribution \cite{Takaki21NatComm}. However, to simplify the fitting procedure we set $\Delta L(F)$ to be the median step-sizes at a given force. The median step size as function of force are shown in Fig. \ref{fig:LEmodel}E(inset).

We need to consider two processes. This first (Process 1) is the $X_0(i)$ to $X_1(i+1)$ step for which the dwell time distribution is given by,
\begin{equation}
   f_1(t) = k_1^+ e^{-k_1^+ t}
\end{equation}
The average dwell time in this case is $1/k_1^+$.

In the second case (Process 2), a
$X_1(i+1)\to X_0(i+1)$ step followed by another $X_0(i+1) \to X_1(i+2)$ step occurs. In this case the dwell time distribution,
\begin{equation} 
\begin{split}
\label{eq:}  f_2(t) &\sim \int_{0}^t dt' k_2^+ e^{-(k_2^+ + k_1^-)t'}k_1^+ e^{-k_1^+ (t-t')} \\
   &= k_1^+ k_2^+\frac{e^{-k_1^+ t}-e^{-(k_1^- + k_2^+)t}}{k_1^--k_1^+ + k_2^+}.
\end{split}
\end{equation}
The above equation is a special case of the Erlang distribution or hypoexponential function.
% In this case the dwell time distribution is 
% \begin{equation}
%     f_2(t) = k_1^+ (k_1^- + k_2^+)\frac{e^{-k_1^+ t}-e^{-(k_1^- + k_2^+)t}}{k_1^--k_1^+ + k_2^+}
% \end{equation}
The dwell time distribution is thus given as $P(\tau_F)\sim P_1 f_1(t)+P_2 f_2(t)$. The mean dwell time  $\langle\tau_F\rangle$ for the forward step is, 
\begin{equation}
  \langle\tau_F\rangle= \int_{0}^\infty dt  (P_1 f_1(t) + P_2 f_2(t))t,
\end{equation}
where $P_1$ is the probability of the first process and $P_2$ is the probability of the second process. Thus, $P_1=k_1^-/(k_2^+ + k_1^-)$ and $P_2=k_2^+ / (k_2^+ + k_1^-)$ respectively. The mean dwell time for the forward step  simplifies to,
\begin{equation}
    \langle\tau_F\rangle=\frac{k_1^-}{k_2^++k_1^-}\frac{1}{k_1^+}+\frac{k_2^+}{k_2^++k_1^-}\frac{\left(k_1^-+k_1^++k_2^+\right)}{k_1^+\left(k_1^-+k_2^+\right)}
\end{equation}\\

Before  $k_2^+$ can be computed, we need to calculate how the length of the incorporated loop changes as a function of force. In our previous work~\cite{Takaki21NatComm}, we derived an approximate form for $P(\vL|\vR,F=0)$ and $P(\vL|\vR,F>0)$ for semi-flexible polymers. $P(\vL\mid \vR,F)$ is the conditional probability  distribution of a semi-flexible polymer with the contour length $\vL$, at the an end-to-end distance $\vR$ under an applied force $F$. In our loop extrusion model, $\vL$ corresponds to the captured loop length $\Delta L$. At $F=0$ this is given as
\begin{equation}
\begin{split}
    P(\vL|\vR) = D_1 \frac{4\pi N\{L\}(\vR/\vL)^2}{\vL(1-(\vR/\vL)^2)^{9/2}}\\
    \cdot\exp\left(-\frac{3t\{\vL\}}{4(1-(\vR/\vL)^2)}\right),
    \label{eq:plr}
\end{split}
\end{equation}
and for $F>0$
\begin{equation}
\begin{split}
    P(\vL|\vR,F>0) = D_2 \frac{4\pi N^2\{L\}(\vR/\vL)^2}{\vL(1-(\vR/\vL)^2)^{9/2}}\\
    \cdot\exp\left(-\frac{3t\{\vL\}}{4(1-(\vR/\vL)^2)}\right)\\
    \cdot\exp\left(\frac{F\vR}{k_BT}-(1.0+3.3e^{F/F_0})\frac{F\vL}{k_BT}\right),
    \label{eq:plrf}
\end{split}
\end{equation}
where $F_0=1/7$ pN, $t\{\vL\}=3\vL/2l_p$, $N\{\vL\}=\left(\frac{4\alpha^{3/2}e^\alpha}{\pi^{3/2}(4+12\alpha^{-1}+15\alpha^{-2})}\right)$ with $\alpha\{\vL\}=3t\{\vL\}/4$. $D_1$ and $D_2$ are different normalization constants that do not depend on $\vL$. These distributions are heavy-tailed with an undefined mean, so we calculate the harmonic mean instead. The harmonic mean is a lower bound for the mean and can be calculated for any probability distribution. For the sake of simplicity we will refer to this harmonic mean as the mean capture length. Assuming $\vR=40$ nm, which is the size of the O-state in condensin and $l_p=50$ nm, which is the persistence length of dsDNA , we numerically compute the mean capture length using Eq. \ref{eq:plrf}. This is shown in Fig. \ref{fig:LEmodel} (E). To obtain the mean capture length at any arbitrary force we can linearly interpolate from this data. 

\subsection{Observation time from experiments}
We assume that the average observation time for each trace generated in experiments is approximately the total number of forward (or reverse) steps multiplied by the average time taken for each forward (or reverse) step. The average number of forward steps and reverse steps taken per trace at $F=0.4$ pN from Fig. 3F in Ryu \textit{et al.} \cite{ryu2022condensin} are approximately 2.4 and 1.02 respectively. The mean time  for each step is the dwell time. We computed the average dwell time from Fig. S4B in Ryu \textit{et al.}~\cite{ryu2022condensin} by summing over the bins in the histogram. This gives an average dwell time of $16.06$ s and $9.95$ s for the forward and reverse steps, respectively. We did not use the average dwell times (12.6 s and 6.6 s) reported in the paper because the dwell time distributions need not be exponential. Thus, at $F=0.4$ pN the total time to observe all the steps is $\tau_F\cdot m_1 + \tau_R\cdot m_2 =$ $16.06\cdot2.4+9.95\cdot1.02=48.69\approx50$s, where $m_1=2.4$ and $m_2=1.02$ are the average number of forward and reverse steps; and $\tau_F=16.06$ s and $\tau_R=9.95$ s are the mean dwell times of the forward and reverse steps, at the condition of $F=0.4$ pN, respectively. The best fit values are listed in Table. \ref{tab:parameters}. 

\vspace{.2in}
\noindent \textbf{Acknowledgments:} This work was supported by the
NSF (grant CHE 2320256) and by the Welch
Foundation (grant F-0019) administered through the Collie-Welch Regents Chair.

\newpage
\bibliography{references}

@article{ryu2022condensin,
  title={Condensin extrudes {DNA} loops in steps up to hundreds of base pairs that are generated by {ATP} binding events},
  author={Ryu, Je-Kyung and Rah, Sang-Hyun and Janissen, Richard and Kerssemakers, Jacob WJ and Bonato, Andrea and Michieletto, Davide and Dekker, Cees},
  journal={Nucleic acids research},
  volume={50},
  number={2},
  pages={820--832},
  year={2022},
  publisher={Oxford University Press}
}

@article{Johnson20PhysBiol,
  title={Dynamic catch-bonding generates the large stall forces of cytoplasmic dynein},
  author={Johnson, Christopher M and Fenn, J Daniel and Brown, Anthony and Jung, P},
  journal={Physical biology},
  volume={17},
  number={4},
  pages={046004},
  year={2020},
  publisher={IOP Publishing}
}

@ARTICLE{Block94Cell,
  author = {K. Svoboda and S. M. Block},
  title = {Force and Velocity Measured for Single Kinesin Molecules},
  journal = {Cell},
  year = {1994},
  volume = {77},
  pages = {773-784},
}

@ARTICLE{Block07BJ,
  author = {S. M. Block},
  title = {Kinesin Motor Mechanics: Binding, Stepping, Tracking, Gating and Limping},
  journal = {Biophys. J.},
  year = {2007},
  volume = {92},
  pages = {2986-2995},
}

@article{Mugnai20RMP,
  title={Theoretical perspectives on biological machines},
  author={Mugnai, Mauro L and Hyeon, Changbong and Hinczewski, Michael and Thirumalai, D},
  journal={Reviews of Modern Physics},
  volume={92},
  number={2},
  pages={025001},
  year={2020},
  publisher={APS}
}

@article{Revyakin06Science,
  title={Abortive initiation and productive initiation by RNA polymerase involve {DNA} scrunching},
  author={Revyakin, Andrey and Liu, Chenyu and Ebright, Richard H and Strick, Terence R},
  journal={Science},
  volume={314},
  number={5802},
  pages={1139--1143},
  year={2006},
  publisher={American Association for the Advancement of Science}
}

@article{Kapanidis06Science,
  title={Initial transcription by {RNA} polymerase proceeds through a {DNA}-scrunching mechanism},
  author={Kapanidis, Achillefs N and Margeat, Emmanuel and Ho, Sam On and Kortkhonjia, Ekaterine and Weiss, Shimon and Ebright, Richard H},
  journal={Science},
  volume={314},
  number={5802},
  pages={1144--1147},
  year={2006},
  publisher={American Association for the Advancement of Science}
}

@article{yildiz2004kinesin,
	Author = {Yildiz, Ahmet and Tomishige, Michio and Vale, Ronald D and Selvin, Paul R},
	Journal = {Science},
	Number = {5658},
	Pages = {676--678},
	Publisher = {American Association for the Advancement of Science},
	Title = {Kinesin walks hand-over-hand},
	Volume = {303},
	Year = {2004}}

@article{Kolomeisky03BJ,
	Author = {Kolomeisky, Anatoly B and Fisher, Michael E},
	Journal = {Biophysical journal},
	Number = {3},
	Pages = {1642--1650},
	Publisher = {Elsevier},
	Title = {A simple kinetic model describes the processivity of myosin-{V}},
	Volume = {84},
	Year = {2003}}

@article{Hinczewski13PNAS,
  title={Design principles governing the motility of {yosin V}},
  author={Hinczewski, Michael and Tehver, Riina and Thirumalai, D},
  journal={Proceedings of the National Academy of Sciences},
  volume={110},
  number={43},
  pages={E4059--E4068},
  year={2013},
  publisher={National Academy of Sciences}
}

@article{Aiipour12NAR,
  title={Self-organization of domain structures by {DNA}-loop-extruding enzymes},
  author={Alipour, Elnaz and Marko, John F},
  journal={Nucleic acids research},
  volume={40},
  number={22},
  pages={11202--11212},
  year={2012},
  publisher={Oxford University Press}
}

@ARTICLE{Marshall03Nature,
  author = {B. T. Marshall and M. Long and J. W. Piper and T. Yago and R. P. {McEver} and C. Zhu},
  title = {Direct observation of catch bonds involving cell-adhesion molecules},
  journal = {Nature},
  year = {2003},
  volume = {423},
  pages = {190-193},
}

@article{Dekker23Science,
  title={How do molecular motors fold the genome?},
  author={Dekker, Cees and Haering, Christian H and Peters, Jan-Michael and Rowland, Benjamin D},
  journal={Science},
  volume={382},
  number={6671},
  pages={646--648},
  year={2023},
  publisher={American Association for the Advancement of Science}
}

@article{Harder15BJ,
  title={Catch bond interaction between cell-surface sulfatase Sulf1 and glycosaminoglycans},
  author={Harder, Alexander and M{\"o}ller, Ann-Kristin and Milz, Fabian and Neuhaus, Phillipp and Walhorn, Volker and Dierks, Thomas and Anselmetti, Dario},
  journal={Biophysical journal},
  volume={108},
  number={7},
  pages={1709--1717},
  year={2015},
  publisher={Elsevier}
}

@article{Bell78Science,
	Author = {Bell, George I},
	Journal = {Science},
	Number = {4342},
	Pages = {618--627},
	Publisher = {American Association for the Advancement of Science},
	Title = {Models for the specific adhesion of cells to cells},
	Volume = {200},
	Year = {1978}}

@article{Luca17Science,
  title={Notch-Jagged complex structure implicates a catch bond in tuning ligand sensitivity},
  author={Luca, Vincent C and Kim, Byoung Choul and Ge, Chenghao and Kakuda, Shinako and Wu, Di and Roein-Peikar, Mehdi and Haltiwanger, Robert S and Zhu, Cheng and Ha, Taekjip and Garcia, K Christopher},
  journal={Science},
  volume={355},
  number={6331},
  pages={1320--1324},
  year={2017},
  publisher={American Association for the Advancement of Science}
}

@article{Huang17Science,
  title={Vinculin forms a directionally asymmetric catch bond with {F}-actin},
  author={Huang, Derek L and Bax, Nicolas A and Buckley, Craig D and Weis, William I and Dunn, Alexander R},
  journal={Science},
  volume={357},
  number={6352},
  pages={703--706},
  year={2017},
  publisher={American Association for the Advancement of Science}
}

@article{Choi25AnnRevImm,
  title={Catch Bonds in Immunology},
  author={Choi, Hyun-Kyu and Zhu, Cheng},
  journal={Annual Review of Immunology},
  volume={43},
  year={2025},
  publisher={Annual Reviews}
}

@ARTICLE{Thomas2008ARB,
  author = {Thomas, W.E. and Vogel, V. and Sokurenko, E.},
  title = {Biophysics of catch bonds},
  journal = {Annu. Rev. Biophys.},
  year = {2008},
  volume = {37},
  pages = {399--416},
  publisher = {Annual Reviews},
}

@article{Fudenberg16CellRep,
  title={Formation of chromosomal domains by loop extrusion},
  author={Fudenberg, Geoffrey and Imakaev, Maxim and Lu, Carolyn and Goloborodko, Anton and Abdennur, Nezar and Mirny, Leonid A},
  journal={Cell reports},
  volume={15},
  number={9},
  pages={2038--2049},
  year={2016},
  publisher={Elsevier}
}

@article{Pobegalov23NatComm,
  title={Single cohesin molecules generate force by two distinct mechanisms},
  author={Pobegalov, Georgii and Chu, Lee-Ya and Peters, Jan-Michael and Molodtsov, Maxim I},
  journal={Nature Communications},
  volume={14},
  number={1},
  pages={3946},
  year={2023},
  publisher={Nature Publishing Group UK London}
}

@article{Kodera10Nature,
	Author = {Kodera, Noriyuki and Yamamoto, Daisuke and Ishikawa, Ryoki and Ando, Toshio},
	Journal = {Nature},
	Number = {7320},
	Pages = {72},
	Publisher = {Nature Publishing Group},
	Title = {Video imaging of walking myosin V by high-speed atomic force microscopy},
	Volume = {468},
	Year = {2010}}

@article{Isojima16NatChemBiol,
	Author = {Isojima, Hiroshi and Iino, Ryota and Niitani, Yamato and Noji, Hiroyuki and Tomishige, Michio},
	Journal = {Nature chemical biology},
	Number = {4},
	Pages = {290},
	Publisher = {Nature Publishing Group},
	Title = {Direct observation of intermediate states during the stepping motion of kinesin-1},
	Volume = {12},
	Year = {2016}}

@article{Dey23CellReports,
  title={Structural changes in chromosomes driven by multiple condensin motors during mitosis},
  author={Dey, Atreya and Shi, Guang and Takaki, Ryota and Thirumalai, D},
  journal={Cell Reports},
  volume={42},
  number={4},
  year={2023},
  publisher={Elsevier}
}

@article{Marko19NAR,
  title={{DNA}-segment-capture model for loop extrusion by structural maintenance of chromosome {(SMC)} protein complexes},
  author={Marko, John F and De Los Rios, Paolo and Barducci, Alessandro and Gruber, Stephan},
  journal={Nucleic acids research},
  volume={47},
  number={13},
  pages={6956--6972},
  year={2019},
  publisher={Oxford University Press}
}

@article{Takaki21NatComm,
  title={Theory and simulations of condensin mediated loop extrusion in DNA},
  author={Takaki, Ryota and Dey, Atreya and Shi, Guang and Thirumalai, D},
  journal={Nature Communications},
  volume={12},
  number={1},
  pages={5865},
  year={2021},
  publisher={Nature Publishing Group UK London}
}

@article{ryu2020condensin,
  title={The condensin holocomplex cycles dynamically between open and collapsed states},
  author={Ryu, Je-Kyung and Katan, Allard J and van der Sluis, Eli O and Wisse, Thomas and de Groot, Ralph and Haering, Christian H and Dekker, Cees},
  journal={Nature structural \& molecular biology},
  volume={27},
  number={12},
  pages={1134--1141},
  year={2020},
  publisher={Nature Publishing Group US New York}
}

@article{ha1995mean,
  title={A mean-field model for semiflexible chains},
  author={Ha, B-Y and Thirumalai, D},
  journal={The Journal of chemical physics},
  volume={103},
  number={21},
  pages={9408--9412},
  year={1995},
  publisher={American Institute of Physics}
}

@article{chakrabarti2017phenomenological,
  title={Phenomenological and microscopic theories for catch bonds},
  author={Chakrabarti, Shaon and Hinczewski, Michael and Thirumalai, D},
  journal={Journal of structural biology},
  volume={197},
  number={1},
  pages={50--56},
  year={2017},
  publisher={Elsevier}
}

@article{Barsegov06JPCB,
  title={Dynamic competition between catch and slip bonds in selectins bound to ligands},
  author={Barsegov, V and Thirumalai, D},
  journal={The Journal of Physical Chemistry B},
  volume={110},
  number={51},
  pages={26403--26412},
  year={2006},
  publisher={ACS Publications}
}

@ARTICLE{Barsegov05PNAS,
  author = {V. Barsegov and D. Thirumalai},
  title = {Dynamics of unbinding of cell adhesion molecules: Transition from catch to slip bonds},
  journal = {Proc. Natl. Acad. Sci. USA},
  year = {2005},
  volume = {102},
  pages = {1835-1839},
}

@article{gillespie1977exact,
  title={Exact stochastic simulation of coupled chemical reactions},
  author={Gillespie, Daniel T},
  journal={The journal of physical chemistry},
  volume={81},
  number={25},
  pages={2340--2361},
  year={1977},
  publisher={ACS Publications}
}

@article{dembo1994lectures,
  title={Lectures on Mathematics in the Life Sciences, Some Mathematical Problems in Biology},
  author={Dembo, M},
  journal={American Mathematical Society, Providence, RI},
  volume={51},
  pages={51--77},
  year={1994}
}

@article{dembo1988reaction,
  title={The reaction-limited kinetics of membrane-to-surface adhesion and detachment},
  author={Dembo, Micah and Torney, DC and Saxman, K and Hammer, D},
  journal={Proceedings of the Royal Society of London. Series B. Biological Sciences},
  volume={234},
  number={1274},
  pages={55--83},
  year={1988},
  publisher={The Royal Society London}
}

@article{Johnson20PhyBiol,
  title={Dynamic catch-bonding generates the large stall forces of cytoplasmic dynein},
  author={Johnson, Christopher M and Fenn, J Daniel and Brown, Anthony and Jung, P},
  journal={Physical biology},
  volume={17},
  number={4},
  pages={046004},
  year={2020},
  publisher={IOP Publishing}
}

@article{Nord17PNAS,
  title={Catch bond drives stator mechanosensitivity in the bacterial flagellar motor},
  author={Nord, Ashley L and Gachon, Emilie and Perez-Carrasco, Ruben and Nirody, Jasmine A and Barducci, Alessandro and Berry, Richard M and Pedaci, Francesco},
  journal={Proceedings of the National Academy of Sciences},
  volume={114},
  number={49},
  pages={12952--12957},
  year={2017},
  publisher={National Academy of Sciences}
}

@article{Barkan24PNAS,
  title={Topology of molecular deformations induces triphasic catch bonding in selectin--ligand bonds},
  author={Barkan, Casey O and Bruinsma, Robijn F},
  journal={Proceedings of the National Academy of Sciences},
  volume={121},
  number={6},
  pages={e2315866121},
  year={2024},
  publisher={National Academy of Sciences}
}

@article{Chakrabarti14PNAS,
  title={Plasticity of hydrogen bond networks regulates mechanochemistry of cell adhesion complexes},
  author={Chakrabarti, Shaon and Hinczewski, Michael and Thirumalai, D},
  journal={Proceedings of the National Academy of Sciences},
  volume={111},
  number={25},
  pages={9048--9053},
  year={2014},
  publisher={National Academy of Sciences}
}

@article{vu2016discrete,
  title={Discrete step sizes of molecular motors lead to bimodal non-Gaussian velocity distributions under force},
  author={Vu, Huong T and Chakrabarti, Shaon and Hinczewski, Michael and Thirumalai, Dave},
  journal={Physical review letters},
  volume={117},
  number={7},
  pages={078101},
  year={2016},
  publisher={APS}
}

@article{ganji2018real,
  title={Real-time imaging of DNA loop extrusion by condensin},
  author={Ganji, Mahipal and Shaltiel, Indra A and Bisht, Shveta and Kim, Eugene and Kalichava, Ana and Haering, Christian H and Dekker, Cees},
  journal={Science},
  volume={360},
  number={6384},
  pages={102--105},
  year={2018},
  publisher={American Association for the Advancement of Science}
}

@article{pobegalov2023single,
  title={Single cohesin molecules generate force by two distinct mechanisms},
  author={Pobegalov, Georgii and Chu, Lee-Ya and Peters, Jan-Michael and Molodtsov, Maxim I},
  journal={Nature Communications},
  volume={14},
  number={1},
  pages={3946},
  year={2023},
  publisher={Nature Publishing Group UK London}
}

@article{uhlmann2016smc,
  title={{SMC} complexes: from {DNA} to chromosomes},
  author={Uhlmann, Frank},
  journal={Nature reviews Molecular cell biology},
  volume={17},
  number={7},
  pages={399--412},
  year={2016},
  publisher={Nature Publishing Group UK London}
}

@article{Strick04CurrBiol,
  title={Real-time detection of single-molecule {DNA} compaction by condensin {I}},
  author={Strick, Terence R and Kawaguchi, Tatsuhiko and Hirano, Tatsuya},
  journal={Current biology},
  volume={14},
  number={10},
  pages={874--880},
  year={2004},
  publisher={Elsevier}
}

@article{nasmyth2005structure,
  title={The structure and function of SMC and kleisin complexes},
  author={Nasmyth, Kim and Haering, Christian H},
  journal={Annu. Rev. Biochem.},
  volume={74},
  number={1},
  pages={595--648},
  year={2005},
  publisher={Annual Reviews}
}

@article{hassler2018towards,
  title={Towards a unified model of SMC complex function},
  author={Hassler, Markus and Shaltiel, Indra A and Haering, Christian H},
  journal={Current Biology},
  volume={28},
  number={21},
  pages={R1266--R1281},
  year={2018},
  publisher={Elsevier}
}

@article{dekker2023molecular,
  title={How do molecular motors fold the genome?},
  author={Dekker, Cees and Haering, Christian H and Peters, Jan-Michael and Rowland, Benjamin D},
  journal={Science},
  volume={382},
  number={6671},
  pages={646--648},
  year={2023},
  publisher={American Association for the Advancement of Science}
}

@article{pradhan2023smc5,
  title={The Smc5/6 complex is a DNA loop-extruding motor},
  author={Pradhan, Biswajit and Kanno, Takaharu and Umeda Igarashi, Miki and Loke, Mun Siong and Baaske, Martin Dieter and Wong, Jan Siu Kei and Jeppsson, Kristian and Bj{\"o}rkegren, Camilla and Kim, Eugene},
  journal={Nature},
  volume={616},
  number={7958},
  pages={843--848},
  year={2023},
  publisher={Nature Publishing Group UK London}
}

@article{davidson2019dna,
  title={{DNA} loop extrusion by human cohesin},
  author={Davidson, Iain F and Bauer, Benedikt and Goetz, Daniela and Tang, Wen and Wutz, Gordana and Peters, Jan-Michael},
  journal={Science},
  volume={366},
  number={6471},
  pages={1338--1345},
  year={2019},
  publisher={American Association for the Advancement of Science}
}

@article{kim2019human,
  title={Human cohesin compacts {DNA} by loop extrusion},
  author={Kim, Yoori and Shi, Zhubing and Zhang, Hongshan and Finkelstein, Ilya J and Yu, Hongtao},
  journal={Science},
  volume={366},
  number={6471},
  pages={1345--1349},
  year={2019},
  publisher={American Association for the Advancement of Science}
}

@article{golfier2020cohesin,
  title={Cohesin and condensin extrude DNA loops in a cell cycle-dependent manner},
  author={Golfier, Stefan and Quail, Thomas and Kimura, Hiroshi and Brugu{\'e}s, Jan},
  journal={Elife},
  volume={9},
  pages={e53885},
  year={2020},
  publisher={eLife Sciences Publications, Ltd}
}

@article{Holzbaur10COCB,
  title={Coordination of molecular motors: from in vitro assays to intracellular dynamics},
  author={Holzbaur, Erika LF and Goldman, Yale E},
  journal={Current opinion in cell biology},
  volume={22},
  number={1},
  pages={4--13},
  year={2010},
  publisher={Elsevier}
}

@article{Svoboda94ARBB,
  title={Biological applications of optical forces},
  author={Svoboda, Karel and Block, Steven M},
  journal={Annual review of biophysics and biomolecular structure},
  volume={23},
  number={1},
  pages={247--285},
  year={1994}
}

@article{kolomeisky2007molecular,
  title={Molecular motors: a theorist's perspective},
  author={Kolomeisky, Anatoly B and Fisher, Michael E},
  journal={Annu. Rev. Phys. Chem.},
  volume={58},
  number={1},
  pages={675--695},
  year={2007},
  publisher={Annual Reviews}
}

@book{doi1988theory,
  title={The theory of polymer dynamics},
  author={Doi, Masao and Edwards, Sam F and Edwards, Samuel Frederick},
  volume={73},
  year={1988},
  publisher={oxford university press}
}

@article{tolman1925principle,
  title={The principle of microscopic reversibility},
  author={Tolman, Richard C},
  journal={Proceedings of the National Academy of Sciences},
  volume={11},
  number={7},
  pages={436--439},
  year={1925}
}

@article{goundaroulisBiophysicalJournal2020,
  title = {Chromatin {{Is Frequently Unknotted}} at the {{Megabase Scale}}},
  author = {Goundaroulis, Dimos and Lieberman Aiden, Erez and Stasiak, Andrzej},
  year = 2020,
  month = may,
  journal = {Biophysical Journal},
  volume = {118},
  number = {9},
  pages = {2268--2279},
  issn = {00063495},
  doi = {10.1016/j.bpj.2019.11.002},
  urldate = {2025-12-29},
  langid = {english}
}

@article{Hirano1997cell,
  title = {Condensins,  Chromosome Condensation Protein Complexes Containing XCAP-C,  XCAP-E and a Xenopus Homolog of the Drosophila Barren Protein},
  volume = {89},
  ISSN = {0092-8674},
  url = {http://dx.doi.org/10.1016/s0092-8674(00)80233-0},
  DOI = {10.1016/s0092-8674(00)80233-0},
  number = {4},
  journal = {Cell},
  publisher = {Elsevier BV},
  author = {Hirano,  Tatsuya and Kobayashi,  Ryuji and Hirano,  Michiko},
  year = {1997},
  month = may,
  pages = {511–521}
}

@article{Michaelis1997cell,
  title = {Cohesins: Chromosomal Proteins that Prevent Premature Separation of Sister Chromatids},
  volume = {91},
  ISSN = {0092-8674},
  url = {http://dx.doi.org/10.1016/s0092-8674(01)80007-6},
  DOI = {10.1016/s0092-8674(01)80007-6},
  number = {1},
  journal = {Cell},
  publisher = {Elsevier BV},
  author = {Michaelis,  Christine and Ciosk,  Rafal and Nasmyth,  Kim},
  year = {1997},
  month = oct,
  pages = {35–45}
}

@article{Brackley2018nucleus,
  title = {Extrusion without a motor: a new take on the loop extrusion model of genome organization},
  volume = {9},
  ISSN = {1949-1042},
  url = {http://dx.doi.org/10.1080/19491034.2017.1421825},
  DOI = {10.1080/19491034.2017.1421825},
  number = {1},
  journal = {Nucleus},
  publisher = {Informa UK Limited},
  author = {Brackley,  C. A. and Johnson,  J. and Michieletto,  D. and Morozov,  A. N. and Nicodemi,  M. and Cook,  P. R. and Marenduzzo,  D.},
  year = {2018},
  month = jan,
  pages = {95–103}
}

@article{Bonato2021biophyjournal,
  title = {Three-dimensional loop extrusion},
  volume = {120},
  ISSN = {0006-3495},
  url = {http://dx.doi.org/10.1016/j.bpj.2021.11.015},
  DOI = {10.1016/j.bpj.2021.11.015},
  number = {24},
  journal = {Biophysical Journal},
  publisher = {Elsevier BV},
  author = {Bonato,  Andrea and Michieletto,  Davide},
  year = {2021},
  month = dec,
  pages = {5544–5552}
}

@article{Yamamoto2017pre,
  title = {Osmotic mechanism of the loop extrusion process},
  volume = {96},
  ISSN = {2470-0053},
  url = {http://dx.doi.org/10.1103/PhysRevE.96.030402},
  DOI = {10.1103/physreve.96.030402},
  number = {3},
  journal = {Physical Review E},
  publisher = {American Physical Society (APS)},
  author = {Yamamoto,  Tetsuya and Schiessel,  Helmut},
  year = {2017},
  month = sep 
}

@article{Higashi2021elife,
  title = {A Brownian ratchet model for DNA loop extrusion by the cohesin complex},
  volume = {10},
  ISSN = {2050-084X},
  url = {http://dx.doi.org/10.7554/eLife.67530},
  DOI = {10.7554/elife.67530},
  journal = {eLife},
  publisher = {eLife Sciences Publications,  Ltd},
  author = {Higashi,  Torahiko L and Pobegalov,  Georgii and Tang,  Minzhe and Molodtsov,  Maxim I and Uhlmann,  Frank},
  year = {2021},
  month = jul 
}

@article{Brackley2017prl,
  title = {Nonequilibrium Chromosome Looping via Molecular Slip Links},
  volume = {119},
  ISSN = {1079-7114},
  url = {http://dx.doi.org/10.1103/PhysRevLett.119.138101},
  DOI = {10.1103/physrevlett.119.138101},
  number = {13},
  journal = {Physical Review Letters},
  publisher = {American Physical Society (APS)},
  author = {Brackley,  C. A. and Johnson,  J. and Michieletto,  D. and Morozov,  A. N. and Nicodemi,  M. and Cook,  P. R. and Marenduzzo,  D.},
  year = {2017},
  month = sep 
}

\end{document}